\documentclass[%
 twocolumn,
superscriptaddress,
 amsmath,amssymb,
 aps, pre
]{revtex4-1}

\pdfoutput=1

\usepackage{graphicx}
\usepackage{dcolumn}
\usepackage{bm}

\usepackage{color}   
\usepackage{commath}
\usepackage{breqn}  
\usepackage{epstopdf}
\usepackage{mathtools} 
\DeclareMathOperator{\erf}{erf}
\DeclareMathOperator{\erfc}{erfc}
\newcommand*\conj[1]{\overline{#1}}
\renewcommand{\figref}[1]{Fig.~\ref{fig:#1}}

\renewcommand{\eqref}[1]{Eq.~(\ref{eq:#1})}

\makeatletter     
\let\cat@comma@active\@empty
\makeatother

\graphicspath{{.},{Fig/}} 

\newcommand{\Ima}{{\rm Im}}

\DeclareGraphicsExtensions{.pdf,.png,.jpg}

\begin{document}


\title{A trophallaxis inspired model for distributed transport between randomly interacting agents}

\author{Johannes Gr{\"a}wer}
 \affiliation{Max Planck Institute for Dynamics and Self-Organization (MPIDS), 37077 G\"ottingen, Germany.}
\author{Henrik Ronellenfitsch}
 \affiliation{Department of Physics and Astronomy, University of Pennsylvania, 19104, Philadelphia, PA}
\author{Marco G. Mazza}
 \affiliation{Max Planck Institute for Dynamics and Self-Organization (MPIDS), 37077 G\"ottingen, Germany.}
\author{Eleni Katifori}
\affiliation{Department of Physics and Astronomy, University of Pennsylvania, 19104, Philadelphia, PA}
 \email{katifori@sas.upenn.edu}

\date{\today}

\begin{abstract}
Trophallaxis, the regurgitation and  mouth to mouth transfer of liquid food between members of eusocial insect societies, is an important process that allows the fast and efficient dissemination of food in the colony. Trophallactic systems are typically treated as a network of agent interactions.
This approach, though valuable, does not easily lend itself to analytic predictions. In this work we consider a simple trophallactic system of randomly interacting agents with finite carrying capacity, and calculate analytically and via a series of simulations the global food intake rate for the whole colony as well as observables describing how uniformly the food is distributed within the nest. Our model and predictions provide a useful benchmark to assess to what level the observed food uptake rates and efficiency in food distribution is due to stochastic effects or specific trophallactic strategies by the ant colony.
Our work also serves as a stepping stone to describing the collective properties of more complex trophallactic systems, such
as those including division of labor between foragers and workers.

\end{abstract}

\pacs{Valid PACS appear here}
\maketitle


\newcommand{\e}{\cdot 10^} 
\newcommand{\dd}{\mbox{d}}

\section{Introduction}

Cooperation and division of labor are the hallmarks of eusocial insect societies such as those of bees, wasps, ants and termites. At the individual level, ants display a rather limited behavioral repertoire~\cite{Holldobler1990}. With a brain as small as one tenth of a cubic millimeter, the individual ant can only make elementary decisions based on local stimuli that confer very small amounts of information~\cite{Razin2013}. However, despite the apparent simplicity of their individual members and the absence of central control, insect societies as a whole exhibit a surprising degree of complexity and can perform complicated tasks such as foraging, food dissemination, brood and nest care that would be inconceivable for a single individual~\cite{Holldobler1990}.

Trophallaxis, the mouth to mouth transfer of liquid food, is considered one of the most central features of eusociality in insects~\cite{Hunt1982}. In ants, it is the main method of liquid food dissemination within the colony. 
In addition, trophallaxis can mediate a uniform colony odor~\cite{Dahbi1999} and confer social immunity against 
pathogens~\cite{Hamilton2011}.
Ants possess two stomachs connected in series, the crop, or storage stomach and the digestion stomach. Forager ants will exit the nest to collect 
food which is temporarily stored in their crop. 
When the forager returns to the nest, she disseminates her crop contents through regurgitation and mouth to mouth feeding to other, non-foraging ants, which in turn disseminate the food further.
Trophallaxis is relevant not only to eusocial insects, but also for other eusocial animals such as bats~\cite{Wilkinson1990}. Beyond its importance for understanding distributed processes in a biological setting, it can provide intuition for engineering applications, such as information or electrical power sharing in swarms of search and rescue robots~\cite{Hamann2007, Schmickl2006, Kubo2004, Ngo2008}.

Because of its importance both in biology and engineering, trophallaxis in general, and in ants in particular, has attracted a lot of attention in the recent years~\cite{Pinter-Wollman2013}. A number of new techniques to precisely measure the movement and location of individuals~\cite{Mersch2013}, the food flow~\cite{Buffin2009,Buffin2012}, or the exact location and food transferred at each trophallactic event simultaneously~\cite{Greenwald2015} have been developed. On the theoretical and computational side, significant progress has been made by using tools from network theory to describe the network of trophallactic interactions in the colony~\cite{gordon2010ant, Waters2012, Blonder2012, Blonder2011a, Fewell2003}.

While this approach has yielded remarkable success, providing 
invaluable insight into the food dissemination process and the 
strategies ants employ to achieve it, the quantitative study of
very basic trophallactic properties is still in its infancy.
Time scales of food distribution and saturation and their relation to parameters such as the colony size are unknown.

In our work, we attempt to fill this gap. We consider a very simple agent-based model of
trophallaxis that captures the main features of the process in real ant colonies, such as finite nest size 
and finite crop capacity, but does not assume any specific ant dissemination strategy, ant identity
(foragers that venture outside the nest versus workers that stay inside), or information transfer between ants. 
Instead, our goal is to understand the collective properties of food dispersion 
in the simplest possible model where the ants inside the nest move at random, pick their partners for food
dissemination at random, and transfer only a specific percentage of their crop content. 
Although we do not consider the movement of the ants explicitly, it is implicitly taken into account in the two models we consider: (\emph{i}) a well mixed colony, where the ants move fast enough to be able to visit and interact with ants at all areas of the nest with equal probability at all times, and (\emph{ii}) a nest with spatial fidelity zones~\cite{Sendova-Franks1995}, where the ants prefer to be localized in small overlapping areas and only interact with other ants that are present in their areas.  

Even in this basic form, the model exhibits a number of interesting properties that 
reproduce some of the behaviors seen in real ant colonies. Due to its simplicity, 
our model allows for analytic predictions in some limits, which provide laws 
of the food dissemination efficiency as a function of the parameters that dictate the
individual ant interactions. These predictions can be viewed as benchmarks to compare 
to the behaviors of real ant colonies, and to assess to what extent the observed 
performance is due to complex strategies and information transfer between the 
donor-receiver pair, and how much they stem from the collective properties of a stochastic system. 

Despite the simplicity of our approach, the two models considered here do not generally reduce to a simple realization of the diffusion equation, 
as one might naively predict. Due to the finite system size and 
nonlinearities introduced by finite crop capacities, they are analytically intractable. 
Since we are interested in the process of transferring food into a nest until saturation,
they are also inherently non-equilibrium. Though we are not interested in the transfer of agents per se  
but in the transfer of food through potentially moving agents, our work is related to some methodologies from 
non-equilibrium transport systems \cite{Schadschneider2010}.

To allow for some analytical predictions, after introducing the agent-based model in 2d, we proceed to focus our attention on 1d systems. We expect the 1d systems to share a lot of the same qualitative behaviors as their 2d counterparts, and to provide useful intuition.

Also, in this work we consider explicitly immobile agents that are only allowed to interact with other agents within a specified interaction radius. 
When this radius is small, the number of potential trophallactic partners is limited. This could also be thought of as ants moving randomly \cite{Sendova-Franks1995, Mersch2013}
around a fixed position in the nest, not allowed to permanently wander away 
from that location. The interaction radius then captures how far they are willing to 
go from their preferred location to interact with a partner. A very large interaction radius is equivalent to a well mixed colony, where ants can quickly visit all of the nest.
A more thorough investigation of the 2d model and finite ant velocity is reserved for future work.

With our work we aim to lay the groundwork for an understanding of stochastic trophallaxis that
will eventually allow to investigate more complex systems with specific strategies in a systematic way.

The rest of this manuscript is structured as follows. In Sec.~\ref{sec:model} we describe in detail the agent based simulation model of trophallaxis  
we will consider in the rest of the work. The simulation model is presented in 2d with explicit motion, 
but the analytical description will be in 1d and without explicit motion. 
In that section, we also present the observables that we will use to describe 
the behavior of the system. In Sec.~\ref{sec:LargeR} we consider the large interaction radius 
limit of our trophallactic model. In this limit every ant can always interact with every other ant in 
the nest, and the analytical description of this agent based model is akin to a mean-field model.  
In Sec.~\ref{sec:stationaryAnts} we discuss to what extent the finite 
interaction radius system can be viewed as a continuous model. We study the food source in
Sec.~\ref{sec:foodSupply} and analytically calculate the characteristic timescale for the food uptake rate of the nest. 
 In Sec.~\ref{sec:Results} we present our numerical results regarding the food uptake rate and distribution within the nest and
in Sec.~\ref{sec:discussion} we finish with a short discussion and summary.

\section{Agent-based simulation model \label{sec:model}}

\begin{figure}
    \includegraphics[width=0.45\textwidth]{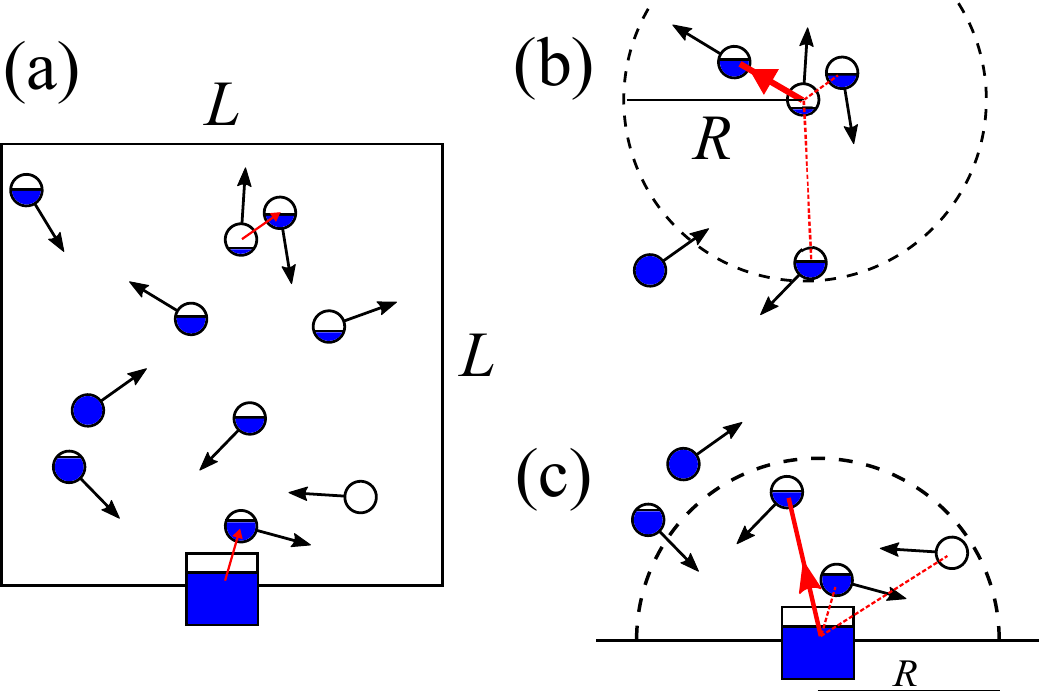}
    \caption{ Graphical representation of the trophallactic process. (a) Sketch of the entire system. The blue color represents the amount of food carried by each agent (circles). An agent is solid blue when filled at capacity ($c_i=c_{\rm{max}}$). The red arrows stand for trophallactic interactions, and the direction of the arrow represents the direction of food transfer. The black arrows signify the direction of agent movement. The source is depicted as a square at the middle of the bottom system wall. (b) Trophallactic interaction between two agents. The donor agent, at the center of the dotted circle, selects one partner at random from the agents inside the interaction radius $R$ (dotted circle) and transfers food. (c) One agent that is within distance $R$ from the source and not yet completely full is selected at random to pick up food from the source. The agent picks up a quantity $(c_{\rm{max}}-c_i)$ from the source to load full.}
    \label{fig:intro_schematic}
\end{figure}

We consider a simple stochastic model of food exchanging, self-propelled agents, confined to a finite nest chamber. 
A graphical representation of the trophallactic process is shown in \figref{intro_schematic}.
The nest (or system) is a square area of size $L \times L$ with a food source located at the center of one boundary, modeling the nest entrance
(see Fig.~\ref{fig:intro_schematic}~(a)).
The $N$ agents (that is, the colony) iteratively perform three basic actions: moving, collecting food from the source, 
and exchanging food with each other. We proceed to describe
the details of these actions in the rest of this section.

\subsection{Agent motion}
We model the agents' motion as active Brownian motion 
of point-like particles (ABPs; see~\cite{Romanczuk2012} for a review) moving with a constant speed $v$ 
and a random unit orientation vector $\hat{e}_{i}$, 
so that the discretized equations of motion for agent $i$ are:
\begin{align*}
    \vec{x}_{i}(t + \Delta t) &= \vec{x}_{i}(t) + \Delta t \cdot v \hat{e}_{i}(t) \\
    \hat{e}_{i}(t + \Delta t) &= \hat{e}_{i}(t) + \Delta t \cdot \left[\eta \hat{\xi_{i}}_{\perp}(t) + \alpha_{i}(t) \hat{e}_{i}(t)\right] ,
\end{align*}
where $\vec{x}_{i}(t) \in [0, L]^{2} \subset \mathbb{R}^{2}$ is the agent's position at time $t$; 
$\Delta t$ is the time step; $\eta$ is a noise parameter, controlling the mean rate of change in orientation; \nolinebreak{
$\hat{\xi_{i}}_{\perp}(t) \coloneqq \hat{\xi_{i}}(t) - \left(\hat{\xi_{i}}(t) \cdot \hat{e}_{i}(t)\right)\hat{e}_{i}(t)$} is the component orthogonal to $\hat{e}_{i}(t)$, of a uniformly distributed random vector on the unit circle $\hat{\xi_{i}}(t)$; and $\alpha_{i}(t)$ is a Lagrangian multiplier, chosen such
that $\abs{\hat{e}_{i}(t)} = 1$ for all times and agents. 
Note that there are no forces between the pointlike agents. We apply hard reflective boundary conditions at the
system boundaries.

This random movement is a gross oversimplification of the complex, occasionally directed motion of ants in 
real colonies.
However, to some extent ants show diffusive mixing behavior in a confined environment like a nest chamber,
such that we believe that an ABP model can serve as a sufficient starting point.

\subsection{Food intake from the source}\label{sec:food_intake}
Each agent $i$ can carry an amount (per agent concentration) of food $c_{i}(t) \leq c_{\rm{max}}$, up to the carrying (or crop) capacity $c_{\rm{max}}$, which is the same for all agents. At $t=0$, no food is inside the system ($c_{i}(0) = 0 \ \forall i$), and enough food to fill every agent ($f_{\rm{max}} = N c_{\rm{max}}$) exists at the food source of the system, which is located at the middle of one boundary ($\vec{x}_{\rm{source}} = (L/2, 0)$). 
The agents have a finite interaction radius $R$ that limits the spatial interaction range, i.e., the distance of the partner with which they can exchange food. Whenever an agent randomly locates the food source, such that the position of the source $\vec{x}_{\rm{source}}$ is within the agent's interaction range ($\abs{\vec{x}_{i}(t) - \vec{x}_{\rm{source}}} \leq R)$), it attempts to pick up food from the source (see Fig.~\ref{fig:intro_schematic}~(c)). Food only enters the system through these uptake events. In order to temporally resolve the food flow into the system, we do not model this food intake events as instantaneous, but consider them to last
a time $T$, called interaction refractory period. 
Therefore, every $T/\Delta t$ time steps, one of the available agents that are within range of the source ($\abs{\vec{x}_{i}(t) - \vec{x}_{\rm{source}}} \leq R)$) and not at their carrying capacity ($c_{i}(t) < c_{\rm{max}}$) is chosen with equal probability, and its food concentration is set to the maximum value $c_{i}(t) = c_{\rm{max}}$. Agents that are at capacity do not attempt to pick up food at the source, so they are ignored even if they are within $R$ of the source. Both the source and the agent that picked up food are then in a refractory state for the next $T/\Delta t$ time steps, in which they are not allowed to exchange food with any other agents. Therefore, agents can only pick up food from the source one at a time. The agent that just picked up food from the source continues to move in its refractory period, such that the agents always keep moving. This is chosen for simplicity, so as to decouple the food exchange events (whether an agent performs an exchange or not) from the random motion. 
It should be noted that real ants can and do feed in parallel from single food sources. This behavior is not 
captured by our model, but for sufficiently small colonies such as those analyzed in~\cite{Greenwald2015},
our description is acceptable.

Note that the exact location of the food source is not important for this model. The model can be modified by moving the source outside the nest, labeling the ants that reach and subsequently exit the nest entrance as {\it foragers} and adjusting the characteristic time between source visits (in our model equal to the refractory period) to account for the extra time needed to reach the source. 

\subsection{Food exchange between agents}
The core of our model is the trophallactic exchange between agents that already carry some food. Little is known about 
the specific details of trophallaxis on the individual level in real animal societies~\cite{Holldobler1990}.
Therefore, we propose a minimal set of interaction rules that reproduces basic trophallaxis dynamics.
Agents that acquired food, randomly choose a food exchange partner within their finite interaction radius $R$
and try to exchange a fixed percentage $\sigma$ of the food that they are carrying (see Fig.~\ref{fig:intro_schematic}~(b)).

The specific rules of the interaction are:
\begin{enumerate}
    \item Every agent $i$ that currently
    \begin{itemize}
        \item has food ($c_{i}(t) > 0$);
        \item is not refractory (its last food exchange or intake from the source was more than $T/\Delta t$ time steps ago);
        \item has at least one other agent $j$ in its interaction range 
        ($\abs{\vec{x}_{i}(t) - \vec{x}_{j}(t)} \leq R$) 
        that is also not refractory
    \end{itemize} 
    is selected in a random order.
    \item The selected agent $i$ randomly chooses one of the non-refractory agents $j$ in its interaction range.
    \item An amount 
    \begin{eqnarray}\label{eq:food_exchange}
        \Delta c_{i \rightarrow j} &=& \min\left(\sigma c_{i}(t), c_{\rm{max}}-c_{j}(t)\right) \nonumber \\
        & = &        \begin{cases}
            \sigma c_{i}(t) & \text{if } \ c_{j}(t) + \sigma c_{i}(t) \leq c_{\rm{max}} \\
            c_{\rm{max}}-c_{j}(t) & \rm{otherwise}
        \end{cases}
    \end{eqnarray} is transferred from agent $i$ to agent $j$. This way, the food receiving agent $j$ takes only as much of the share $\sigma c_{i}(t)$ from the donating agent $i$ as it can carry.  
    In the special case of $c_{j}(t) = c_{\rm{max}}$, $\Delta c_{i \rightarrow j} = 0$, and no food is transferred.
    \item Both food exchange partners $i$ and $j$ are immediately set to be refractory for the next $T/\Delta t$ time steps, irrespective of the actual amount $\Delta c_{i \rightarrow j} \geq 0$ transferred. In consequence, both agents cannot give or receive food again in this iteration (and the next $T/\Delta t$ time steps as well).
\end{enumerate}
This set of rules introduces no bias in the random choice of available food exchange partners and
requires no active information exchange between agents. The single agent always offers the percentage $\sigma$ of 
its own food to a randomly chosen partner, without knowing if the other one is already full or completely empty. It
can only infer that its partner is at the carrying capacity after the food exchange when $\Delta c_{i \rightarrow j} < \sigma c_{i}(t)$.
It is important to note that consequently both the motion, and the trophallactic strategy (food exchange rules) between the
agents are random.

\subsection{Observables}
In this work we are interested in understanding how food, initially localized at the source, spreads through the
system, until every agent is at its carrying capacity, i.e. until $c_{i}(t) = c_{\rm{max}}$ $\forall i$.
Quantifying this non-equilibrium transport process is complex, and care needs to be taken in choosing appropriate
observables. The simplest quantities of interest we will use as observables are the mean concentration of food in the nest 
\begin{equation}
\langle c(t) \rangle = \frac{1}{N}\sum_{i=1}^N c_i(t)
\end{equation}
and the variance of food concentrations
\begin{equation}
\langle \Delta c^2 \rangle = \frac{1}{N}\sum_{i=1}^N \left[c_i(t) - \langle c(t) \rangle\right]^2.
\end{equation}
These observables are not directly informative about the spatial distribution of the food that has entered the nest. 
In order to quantify this, we define the mean squared distance $\rm{MSD}(t)$: 
\begin{align*}
    \text{MSD}(t) \coloneqq \langle d_{i}(t)^{2} c_{i}(t)\rangle_{i} = \frac{1}{N} \sum_{i=1}^{N} d_{i}(t)^{2} c_{i}(t) , \\
\end{align*}
where $d_{i}(t) \coloneqq \abs{\vec{x}_{i}(t)-\vec{x}_{\rm{source}}}$ is the distance between agent $i$ and the food source at time $t$. This quantity is a generalization of the mean squared displacement of a Brownian particle in statistical mechanics. In our case, the distance of each agent from the reference point is weighted by the amount of food it is carrying.

Since $\lim_{t \to \infty} c_{i}(t) = c_{\rm{max}}$ and the system is finite, we can easily calculate the ensemble
average steady state value of $\text{MSD}(t)$ from the system geometry by considering
 \begin{equation}
 \lim_{\substack{t \to \infty\\N \to \infty}} \frac{1}{N} \sum_{i=1}^{N} d_{i}(t)^{2} c_{i}(t) = c_{\rm{max}} \frac{1}{A} \int_A (\vec{x}-\vec{x}_{\rm{source}})^2 \dif A.
 \end{equation} 
 Here, $A$ is the total system area. 
 For a 2d square nest of side length $L$ and the source at the midpoint of one of the sides 
 $\lim_{\substack{t \to \infty\\N \to \infty}} \text{MSD}(t) = \frac{5}{12} c_{\rm{max}} L^{2}$, whereas for 
 a 1d interval nest of length $L$ with the source at one of the endpoints it is simply $\lim_{\substack{t \to \infty\\N \to \infty}} \text{MSD}(t) = \frac{1}{3} c_{\rm{max}} L^{2}$.

We can then define a dimensionless version of $\rm{MSD}(t)$ which we designate by $\rm{\overline{MSD}}(t)$:
\begin{align*}
    \mathrm{\overline{MSD}}(t) \coloneqq \frac{\mathrm{MSD}(t)}{\lim_{\substack{t \to \infty\\N \to \infty}} \mathrm{MSD}(t)} .
\end{align*}
With this observable, we are able to monitor the spatio-temporal distribution of food from the source through the system. Note that $\rm{\overline{MSD}}(t) \in [0, 1]$ and when $\rm{\overline{MSD}}(t) = 1$ the steady state has been reached (colony is full).

\section{Mean-field limit: describing a well mixed nest}\label{sec:LargeR}

To understand the limits of trophallactic behavior in our system, we will first consider the simple well mixed case that takes place when the interaction radius is $R \ge L_{\rm{max}}$, where $L_{\rm{max}}$ is the maximum distance between any two points in the nest, or equivalently when the velocity of the agents becomes  very large, $v \gg L/\Delta t $. In this case, every agent has the chance to interact with every other agent and the source at all times. 
We can formulate an analytic mean-field model if we consider the food transfer as a continuous process. This condition can be satisfied if the food exchange rate is small compared to the crop capacity $c_{\rm{max}}$. This condition in principle requires $\sigma \ll 1$, but also that the ants pick up an infinitesimally small quantity of food at the source. The second constraint can never materialize within the confines of our model. However, we will see that the mean-field model can accurately predict the average food intake of the nest even if both of these requirements are not met.

During a short time interval $[t,t+\dif t]$ each ant $i$ 
interacts with another ant $j$ with probability 
$\dif t/\left[T(N-1)\right]$,
where $\dif t/T$ is the probability of not being in the refractory state
and $1/(N-1)$ is the probability of picking ant $j$ for
interaction.
The amount of food that is transferred from ant $i$ to ant $j$
is given by $\Delta c_{i\rightarrow j}$ of \eqref{food_exchange}.
In addition, each ant picks up food from the source with the probability
$\left[\dif t(c_{\rm{max}}-c_i)\right]/(T_sN)$. The quantity $T_s$ is the average time between visits of ants to the nest entrance.

Thus, the food dynamics is governed by the ordinary differential equation
\begin{eqnarray}
 \label{eq:model}
 \frac{\dif c_{i}(t)}{\dif t}  &=& \frac{1}{T(N-1)}  \sum_{\stackrel{j=1}{j \neq i}}^{N}  \left(\Delta c_{j\rightarrow i} 
 - \Delta c_{i\rightarrow j} \right)
     \nonumber \\
     &&+  \frac{1}{T_s N} (c_{\rm{max}}-c_i).
\end{eqnarray}

This equation can easily be solved analytically for the average food concentration $\langle c(t)\rangle$. 
When taking the average, the summations in \eqref{model} with 
piece-wise defined functions that are symmetric with respect to $i$ and $j$ cancel and the solution finally reads: 
\begin{equation}
    \label{eq:mean-field-food-mean}
    \langle c(t) \rangle = c_{\rm{max}} \left(1-e^{-\frac{t}{NT_s}} \right) .
\end{equation}

In this work, for simplicity, we assume $T_s=T$, as described in Sec.~\ref{sec:food_intake}.
A similar equation for the number of fed individuals as a function of time was derived in \cite{Sendova-Franks2010}, and an exponentially saturating uptake of food has also been observed experimentally in \cite{Greenwald2015}.

\begin{figure}
    \centering
    \includegraphics[width=0.45\textwidth]{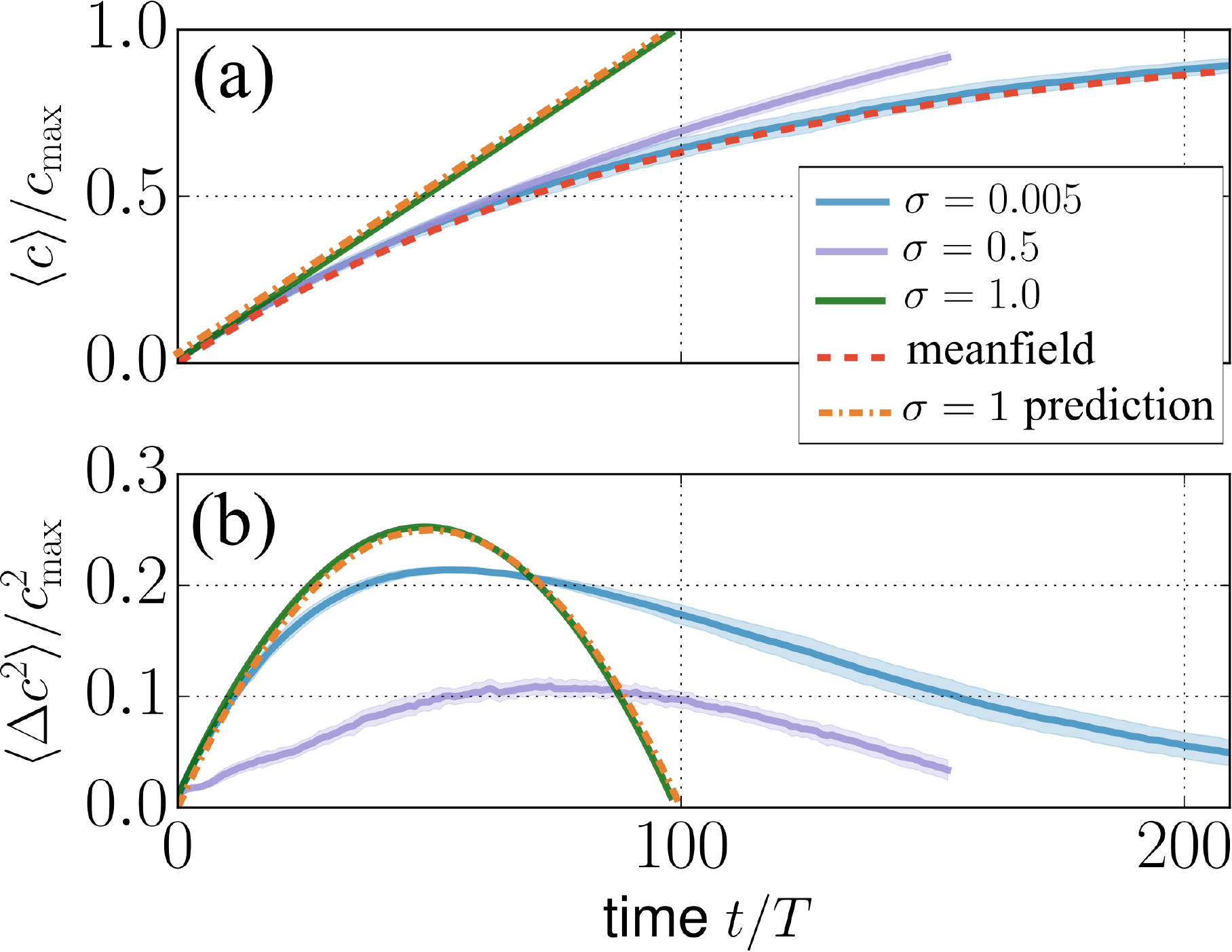} 
    \caption{(a) Average food concentration $\langle c \rangle$ (measured in units of $c_{\rm{max}}$) and (b) food variance $\langle \Delta c^2 \rangle$ (measured in units of $c_{\rm{max}}^2$) versus time (in units of $T$) when $R = L_{\rm{max}}$.  
    Green solid line: $\sigma=1$, purple solid line:  $\sigma=0.5$, blue solid line  $\sigma=0.005$, red dashed line: mean-field prediction (\eqref{mean-field-food-mean}), orange dash-dotted line: $\sigma=1$ prediction (\eqref{meanfield-sigma1-food-mean} for (a) and \eqref{meanfield-sigma1-food-variance} for (b)). Even for $\sigma$ as high as $0.5$, the mean-field model offers a very good approximation of the food uptake dynamics. The width of the shaded area around each line indicates the standard deviation. 
      \label{fig:food-mean-and-variance-share}}
\end{figure}

Figure \ref{fig:food-mean-and-variance-share}  directly compares  the analytic mean-field prediction for the average food concentration of \eqref{mean-field-food-mean} (red dashed line),  with the ensemble averaged simulation data without any fitting parameters for $\sigma=0.005$ (blue line), $\sigma=0.5$ (purple line), and $\sigma=1$ (green line). 
The interaction radius was set to its maximum value $R=L_{\rm{max}}(=L$~in~1d), so that the non-dimensional parameter $\lambda \coloneqq \frac{R}{L}$ (in the rest of the paper called interaction parameter) is $\lambda=1$. The number of ants $N$ was set to $N=100$. For small values of $\sigma$, \eqref{mean-field-food-mean} gives an excellent prediction of the food uptake from the agent-based simulations. As $\sigma$ increases and especially for values
$\sigma \gtrsim 0.5$, the prediction becomes progressively worse. In fact, the mean-field theory is valid for a certain initial time interval that decreases with increasing food exchange ratio $\sigma$. For a low value ($\sigma = 0.005$), agreement is achieved throughout the whole simulation. 

In our simulations ants do not attempt to pick up food from the source when their crop is full. This is in disagreement with the set-up of the mean-field model as the last term of \eqref{model} is normalized with the total number of ants, and not with the number of empty ants. Modifying the feeding rules to allow ants to attempt to pick up food from the source even when full only minimally affects the simulation results and the agreement with the mean-field model is still very good. 

Early in the food dissemination process, the majority of the ants have a completely empty crop. For small times, we can initially neglect the food exchange between ants as the food distribution in the nest is mostly dominated by food uptake events from the entrance. The population of ants is then roughly separated into two groups: the ants that have a full crop, whose number $n$ increases linearly with time as $n=t/T$; and the ants who are empty. Consequently, the total food in the nest increases linearly with time as 
\begin{align}
	\label{eq:meanfield-sigma1-food-mean}
	\langle c\rangle\approx \frac{c_{\rm{max}} t}{NT}
\end{align}    
(orange dot-dashed line in Fig.~\ref{fig:food-mean-and-variance-share}(a)). Note that this binary model becomes exact when $\sigma=1$, as in this case, the ants exchange the entirety of their crop contents and the separation of the ant population in two groups, full and empty, is always exact. 

The variance $\langle \Delta c^2 \rangle$ cannot be predicted by the mean-field model with the same ease. 
However, again for small times, \eqref{meanfield-sigma1-food-mean} leads to
\begin{align}
	\label{eq:meanfield-sigma1-food-variance}
    \langle \Delta c^2 \rangle \approx c_{\rm{max}}^2 \frac{t}{NT} \left(1 - \frac{t}{NT} \right)
\end{align}
(orange dot-dashed line in Fig.~\ref{fig:food-mean-and-variance-share}(b)), which becomes exact for $\sigma=1$ (green line in Fig.~\ref{fig:food-mean-and-variance-share}(b)).

\begin{figure}
    \centering
    \includegraphics[width=0.45\textwidth]{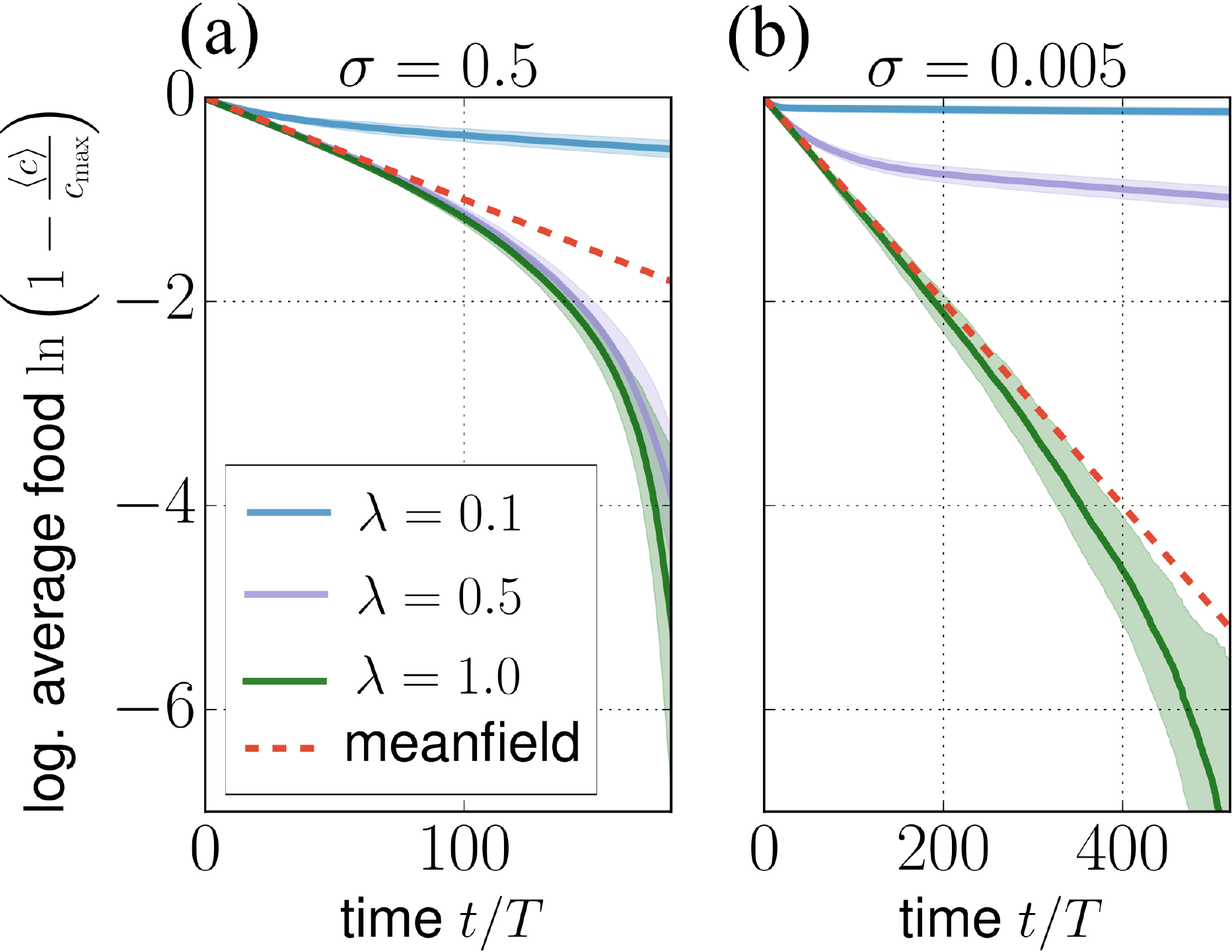} 
    \caption{Average food as a function of time. At the beginning of the
    dynamics, the amount of food in the system is described well by the mean-field model (red dashed line). Later, different dynamics occur, depending on the parameter values.
    (a) For a large $\sigma$ ($\sigma=0.5$), the amount of food is well described by the mean-field model for values of
    $\lambda$ between 1 and 0.5.
    (b) For a small $\sigma$ ($\sigma=0.005$), the mean-field model is valid for a reasonable length of time only for $\lambda\rightarrow 1.0$. The system switches from a fast food uptake rate that agrees very well with the large $R$ model in the initial stages of the process to a slower dynamics that depends on the interaction radius. The time when the transition
occurs decreases with the interaction radius.
    The width of the shaded area around each line indicates the standard deviation.
    \label{fig:food-mean-share-finiteR}}
\end{figure}

Eventually, assuming large velocities $v$ or large radii $R$, the mean-field model predicts that the food is taken up from the source at an exponentially decreasing rate, with decay constant $\gamma_{\lambda=1} = 1/NT$ to a very good approximation. In Fig.~\ref{fig:food-mean-share-finiteR} we explore to what extent this finding holds when $R < L$ (or $\lambda~<~1$), that is, the interaction radius is only a fraction of the nest size. In particular, we consider the food uptake for immobile agents and $\lambda=0.1,\; 0.5$  and $1$ for $\sigma=0.5$  (Fig.~\ref{fig:food-mean-share-finiteR}(a)) and $\sigma=0.005$ (Fig.~\ref{fig:food-mean-share-finiteR}(b)). 

For small $\sigma$ we observe that the system switches from a fast food uptake rate that agrees very well with the large $R$ model in the initial stages of the process to a slower dynamics that depends on the interaction radius. The time when the transition
occurs decreases with the interaction radius. 
There is a simple interpretation for this behavior. Initially, all the ants near the nest entrance that have access to the source are empty, and the trophallactic process dynamics is dominated by the ants at the entrance picking up food. The mean-field theory holds until 
the moment when the ants that can pick up food from the source are approximately at capacity. After this transition, the behavior stops being well explained by mean-field theory
because the food is now diffusing out of the nest entrance vicinity via ant to ant trophallactic interactions.
The transition between the two regimes is abrupt for small $\sigma$, 
and becomes less pronounced for large $\sigma$.

Past the transition time, the food uptake can be approximated by
\begin{equation}
\langle c(t) \rangle=c_{\rm{max}}-\left(c_{\rm{max}}-\langle c(t_0) \rangle\right) e^{-\gamma (t-t_0)},
\label{eq:food-decay-after-transition}
\end{equation}
where $t_0$ is the transition time and $\gamma$ is a new inverse timescale for the food uptake.
The time $t_0$ of the transition increases with $\lambda$.

Sections \ref{sec:stationaryAnts} and \ref{sec:foodSupply} will be devoted to deriving an approximate value for
$\gamma$. We will construct a continuum model to approximate how the food diffuses away from the nest entrance vicinity to the rest of the nest
and use this to study the saturation behavior of food in the colony.

\section{The stationary limit: A nest with spatial fidelity zones} \label{sec:stationaryAnts}

In the previous section, we mainly discussed a mean-field type model that allowed spatially independent, all-to-all interactions. 
In this section, we explore the regime of small interaction radii.
As already mentioned, this can be interpreted as ants that occupy spatial fidelity zones.
In this limit, the behavior of the system becomes strongly spatially dependent. The goal of this section is to derive a set of equations that describe the food dissemination process in this case.

Given the nature of the trophallactic process with $R\ll L$, it is tempting to describe it as a simple diffusive process with a fixed diffusion coefficient and a source term. This is deceptive, as the finite crop capacities, the finite system size, and agents picking up food to capacity at the source render this problem more complex. 
All these conditions are relevant to trophallaxis, but make the problem of finding an exact solution to the appropriate dynamical equations analytically intractable. 
In order to gain some intuition about the behavior of the system, we will adopt a series of simplifying assumptions that allow us to analytically solve the equations, derive a trophallactic timescale and demonstrate that the overall system behavior qualitatively agrees with the full trophallactic dynamics, obtained by simulations. 

While the mean-field model was spatially independent and its predictions should hold for any number of dimensions, any spatially dependent model will in principle depend on the dimensionality of the nest.
Since we are primarily seeking to understand qualitative behaviors, the continuous limit trophallactic models in this work will be written down and solved in 1d. An explicit treatment of the more realistic 2d model will be presented in a future publication.

\begin{figure}
    \includegraphics[width=0.5\textwidth]{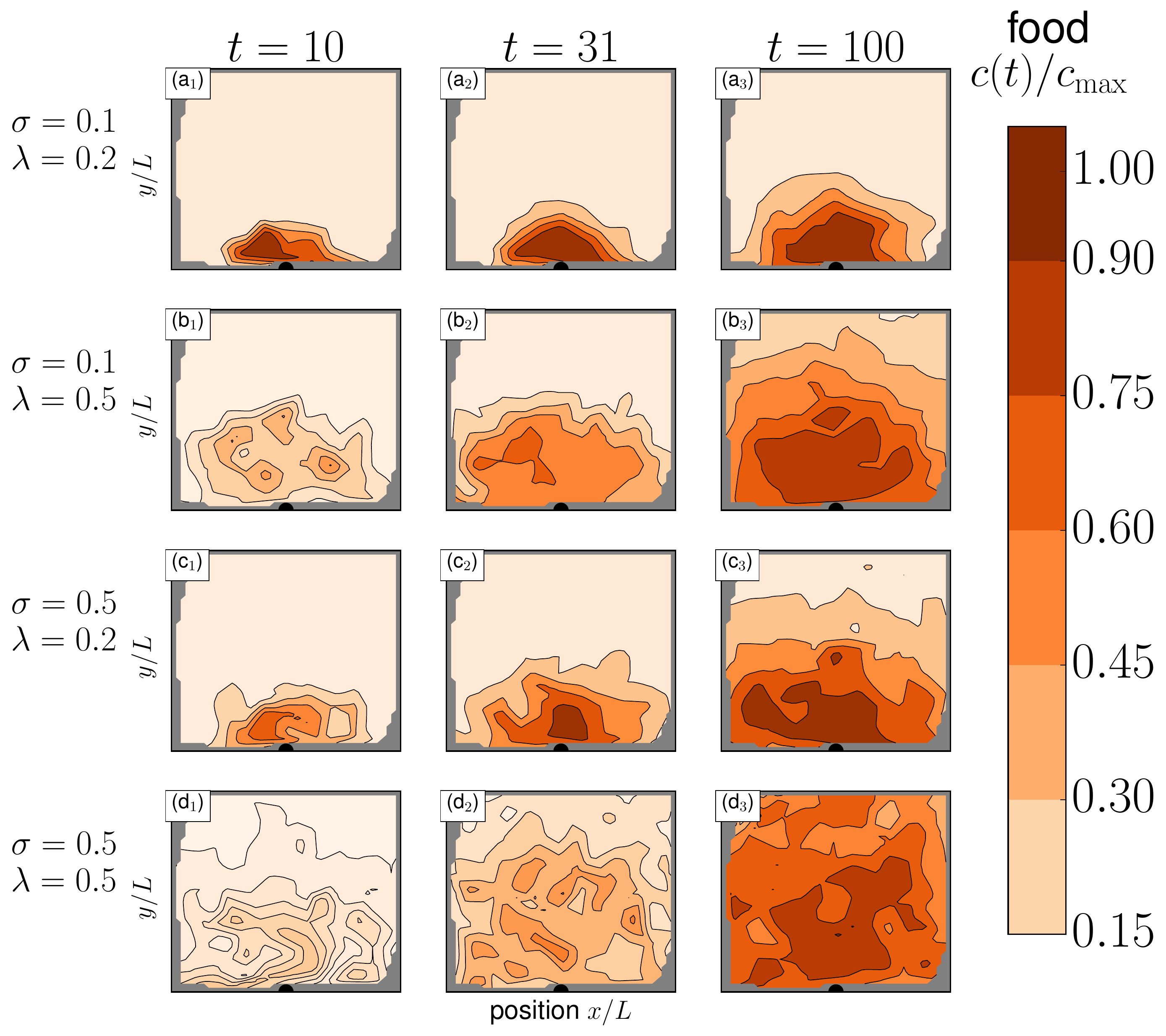}
    \caption{Time series of 2d spatial food distribution pattern for various parameter values $\sigma$ and $\lambda$.
    For small values of $\sigma$ and $\lambda$, the dynamics is approximately diffusive with food
    being distributed locally amongst close agents. For larger values,
    food becomes delocalized and spread among many agents that are far from each other.
    \label{fig:food_spatial_distribution_2}}
\end{figure}

\begin{figure}
    \includegraphics[width=0.495\textwidth]{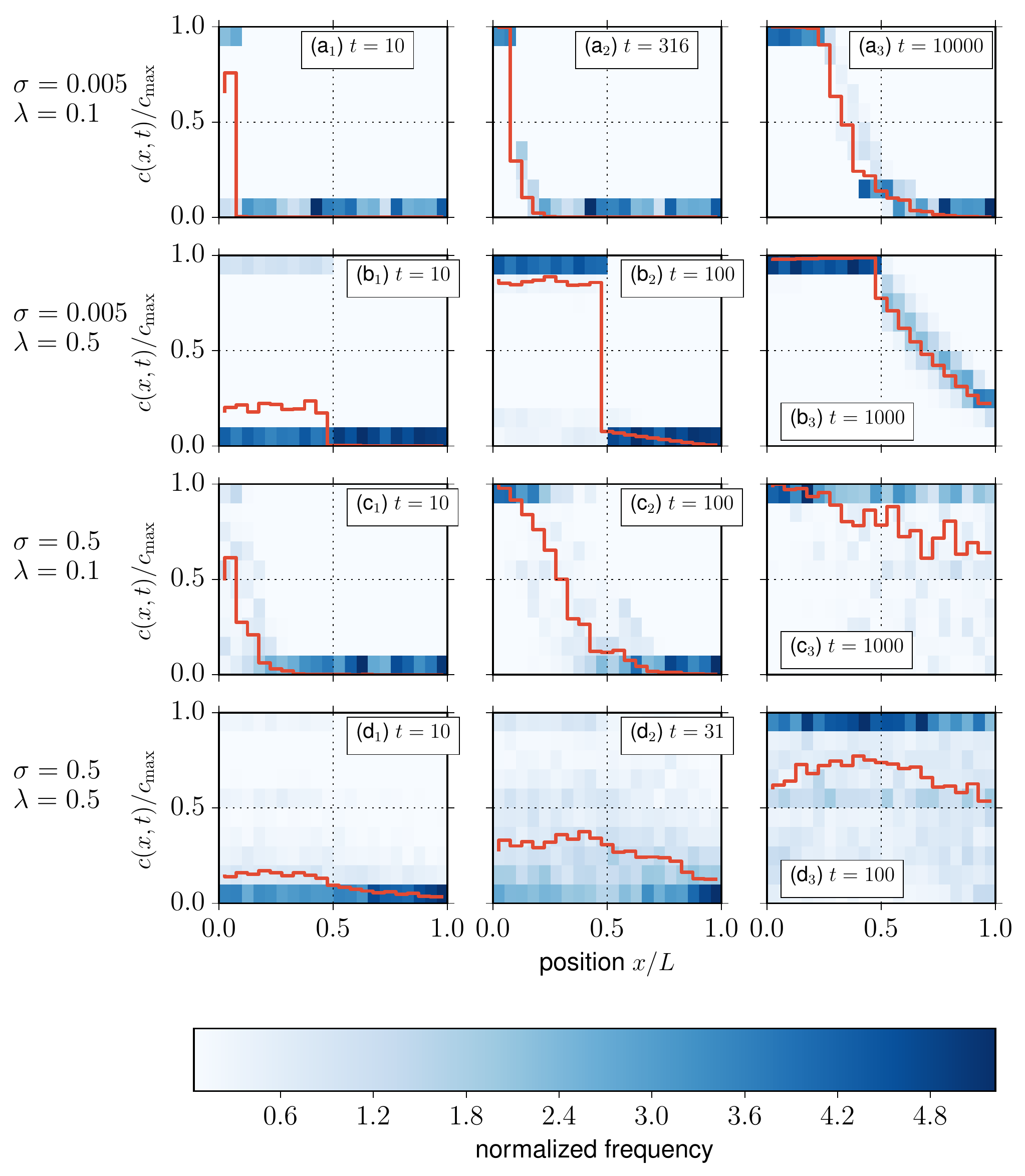}   
    \caption{Food distribution patterns for ants in 1d. The red line is an ensemble average over 100 realizations of
    the model, the blue background is a histogram of the food distribution among the ants at each point in space. The 1d behavior and concentration dependence on $\sigma$ and $\lambda$ qualitatively follows that of the 2d system, displayed in Fig.~\ref{fig:food_spatial_distribution_2}.}
    \label{fig:food_spatial_distribution_1}
\end{figure}

As we see by comparing Fig.~\ref{fig:food_spatial_distribution_2} and  Fig.~\ref{fig:food_spatial_distribution_1}, the behaviors of the 1d and 2d system are qualitatively similar. In Fig.~\ref{fig:food_spatial_distribution_2}, we present a time series of the food distribution in a square nest for various parameters $\sigma$ and $\lambda$. The food source of the nest is at the center of the bottom nest border, and indicated with a black semicircle. 
One hundred ants are initially distributed at random in the nest, with a uniform probability distribution. Since the uniform probability typically produces a very inhomogeneous distribution of ants in the nest, a repulsive potential is applied to ensure that the ants occupy the nest more homogeneously before their position is fixed.  
In Fig.~\ref{fig:food_spatial_distribution_2} we present the food distribution averaged over 10 realizations of the simulation for the same ant positions. 
In Fig.~\ref{fig:food_spatial_distribution_1} we show the average food distribution (solid red line) for for various parameters $\sigma$ and $\lambda$, when the ants are confined to 1d. Figures \ref{fig:food_spatial_distribution_2} and \ref{fig:food_spatial_distribution_1} indicate several qualitative similarities between the 2d and 1d system. 
First, for small $\sigma$ and $\lambda$ (row (a)), the food initially quickly saturates the area at the vicinity of the nest entrance (the area within interaction range of the source), and then disperses to the whole nest via an initially narrow and subsequently broadening moving front. 
Second, for small $\sigma$ but larger $\lambda$ (row (b)), the area within interaction range of the source takes longer to saturate before the food diffuses away. 
Third, for large $\sigma$ and small $\lambda$ (row (c)) there is no clear separation between the time the area close to the nest entrance fills at capacity and the time the diffusion of food out of that area becomes important. However, like in the previous cases, there is still a gradient of food density from the source at the nest entrance to the opposite border of the nest. 
Finally, for large $\sigma$ and $\lambda$ (row (d)), the gradient disappears, and the nest acquires food roughly uniformly.

In order to write down models that are analytically solvable, we will replace the discrete, agent based description with a formalism that is continuous in space and food exchange. In this continuous limit, it is more informative if we speak in terms of spatial food densities $\rho(x, t)$ instead of per-agent concentrations $c(x_i, t)$.
The relationship between the continuum and discrete food variables
at a position $x$ and time $t$ is given 
by multiplying the concentration with the ant number density $N/L$:
\begin{align}
\label{eq:rho-c-relation}
    \rho(x, t) \coloneqq c(x, t) \frac{N}{L} \ .
\end{align}
The maximal amount of food $f_{\rm{max}}$ in the system is then related to the concentration carrying capacity $c_{\rm{max}}$ and the density carrying capacity $\rho_{\rm{max}} = \lim_{t \to \infty } \rho(x, t) $ as 
\begin{align}
\label{eq:rho-max}
    f_{\rm{max}} = L \rho_{\rm{max}} = N c_{\rm{max}} .
\end{align}

For simplicity, we initially assume
that the ants, or points in space, only interact with their nearest neighbors, i.e., we consider the case of
small $\lambda$. 
The balance equation describing the food exchange between neighboring points away from the source at positions
$x$, $x+\Delta x$, $x-\Delta x$ is 
\begin{align}
    \label{eq:balance}
    \Delta t \frac{\partial \rho(x, t)}{\partial t} &=
    \Delta\rho_{x+\Delta x\rightarrow x} - \Delta\rho_{x\rightarrow x+\Delta x} \\
    \nonumber
    &- \Delta\rho_{x\rightarrow x - \Delta x} + \Delta\rho_{x-\Delta x\rightarrow x} \ ,
\end{align}
where $\Delta\rho$ is the continuous analogue to \eqref{food_exchange}. Using \eqref{rho-c-relation}, this is:
\begin{eqnarray*}
    \Delta\rho_{a\rightarrow b} &=& \begin{cases}
        \sigma \rho(a) & \text{if } \sigma \rho(a) + \rho(b) \leq \rho_{\rm{max}} \\
        \rho_{\rm{max}} - \rho(b) & \text{otherwise}
    \end{cases} \\
    &=& \sigma \rho(a) \Theta\left(\frac{\rho_{\rm{max}} - \sigma \rho(a) - \rho(b)}{\rho_{\rm{max}}}\right) \\
    & &\qquad+ (\rho_{\rm{max}}- \rho(b)) \Theta\left(\frac{\sigma \rho(a) + \rho(b) - \rho_{\rm{max}}}{\rho_{\rm{max}}}\right) \ ,
\end{eqnarray*}
where $\Theta(x)$ is the Heaviside step function.
By replacing $\rho(x,t)$ with its Taylor expansion in $\Delta x$ up to the order $O(\Delta x^2)$, we can obtain a diffusion-like spatio-temporal partial differential equation for $\rho(x,t)$:
\begin{align*}
    \rho(x \pm \Delta x) = \rho(x) \pm \Delta x \frac{\partial \rho}{\partial x} + \frac{1}{2}
    \Delta x^2 \frac{\partial^2 \rho}{\partial^2 x} \ .
\end{align*}
Since the Heaviside functions 
are not differentiable, we regularize them using the identity
\begin{align*}
    \Theta(x) = \lim_{k\rightarrow\infty} \Theta_k(x)
    = \lim_{k\rightarrow\infty} \left( \frac{1}{2} + \frac{1}{\pi}
    \arctan(k x) \right).
\end{align*}
By first replacing all $\Theta(x)$ by $\Theta_k(x)$, then expanding $\rho(x,t)$, and finally taking the limit 
$k\rightarrow\infty$, we derive
\begin{align} 
\label{eq:foodDiffusionCropCapacity}
    \frac{\partial \rho}{\partial t} &= \begin{cases}        \sigma \frac{\Delta x^{2}}{\Delta t}  \frac{\partial^2 \rho}{\partial^2 x} & \text{if } \frac{\rho}{\rho_{\rm{max}}} (1+\sigma) < 1 \\   
\frac{\Delta x^{2}}{\Delta t}  \frac{\partial^2 \rho}{\partial^2 x} &  \text{if } \frac{\rho}{\rho_{\rm{max}}} (1 + \sigma) > 1. \end{cases}
\end{align}
If ants are evenly spaced and each ant interacts only with its nearest neighbors, then $\Delta x \approx L/N$ and $\Delta t=T$. 
If the ants are not evenly spaced, but each ant still interacts only with its nearest neighbors, we can modify \eqref{balance} to account for the variability in $\Delta x$ and then perform an ensemble average over all agent distributions. Finally, \eqref{foodDiffusionCropCapacity} is modified by substituting $\Delta x^2$ with  $\langle \Delta x^2 \rangle$, the ensemble mean squared distance between the particles. More generally, if the finite interaction radius allows for an agent to have more than one potential trophallactic partner, then $\langle \Delta x^2 \rangle$ will be replaced by $\overline{r^2}$, the ensemble averaged squared distance between an ant and all its possible interaction partners.
Assuming randomly distributed ants on the one dimensional interval $x \in [0,L]$, 
this average distance can be analytically calculated in the continuous limit. The resulting expression $\overline{r^{2}}$ is a function of the interaction radius $R$.
We show the full calculation in Appendix~\ref{sect:interaction-radius}.
The final result is
\begin{align}
    \frac{\overline{r^{2}}}{L^{2}} = \frac{1}{3}\begin{cases} -\frac{1}{3}\lambda^{3}+\lambda^{2} & \text{if } \lambda \in [0,\frac{1}{2}) \\
    -\frac{5}{3}\lambda^{3}+3\lambda^{2}-\lambda+\frac{1}{6} & \text{if } \lambda \in [\frac{1}{2},1].
    \end{cases}
   \label{eq:interaction-radius}
\end{align}
\eqref{foodDiffusionCropCapacity} is a diffusion equation with a density dependent diffusion constant. For  densities smaller than a threshold $\rho_{\rm{max}}/(1+\sigma)$ the effective diffusion constant is $D=\sigma\overline{r^{2}}/T$, whereas for densities larger than that threshold the diffusion constant increases to $D=\overline{r^{2}}/T$. 
Note that as a consequence of finite crop capacity, very close to the nest entrance food diffuses effectively faster and not slower, as one might naively expect.

In the next section we will show that the timescale derived from the below threshold regime determines the evolution of the system.

\section{Diffusive limit: Describing the dynamics of food uptake}\label{sec:foodSupply}

In the previous section, we showed an estimate of the effective diffusion constant in the nest, and saw that the diffusion constant depends on the 
food concentration in the nest (see \eqref{foodDiffusionCropCapacity}). 
Solving this diffusion equation analytically for any boundary conditions or source term is not straightforward.
We assume that in the initial stages of the trophallactic process the majority of the agents far from the source are below carrying capacity, (in particular that $\rho(x,t)<\rho_{\rm{max}}/(1+\sigma)$), so the diffusion constant is 
\begin{equation}\label{eq:diffusionConstant}
D=\sigma \overline{r^2}/T. 
\end{equation}
In reality, the diffusion constant should switch to its high food density value near the source (see \eqref{foodDiffusionCropCapacity}), but for simplicity, we will just consider $D=\rm{const}$ in the following derivations of this section.
In addition, it is a priori unclear how to handle the food source. 
At first sight, the food supply of the nest could be treated as an initial condition of the diffusion equation, e.g. a delta distribution located at the nest entrance. However, this would turn the nest entrance into yet another food carrying point in space, indistinguishable from the ants. But that is not the case, since the nest entrance cannot receive food and does not have a carrying capacity. The food located at the nest entrance is not really part of the system, which is why it should rather be treated either as a boundary condition or as a source term of the diffusion equation.
We explore these two possibilities in the rest of this section. 
First, we model the food source as a boundary condition, fixing the amount of food at the nest entrance, and 
assuming a fixed diffusion coefficient.
Second, we model the source as an explicit source term, continuously disbursing food into the nest and in addition consider
finite crop capacity.
In both cases we derive characteristic saturation time scales and show they are in agreement. 

\subsection{The food source as a boundary condition}
\label{sec:source-as-boundary}

At the source, the agents should be at carrying capacity. This means that the source can be implemented as a boundary condition. Namely, if the source is at $x=0$, then we fix $\rho(0,t)=\rho_{\rm{max}}$. 
We therefore have to solve the standard diffusion equation without a source term in a 1d finite system.
The appropriate boundary and initial conditions are
\begin{align*}
    \rho(x,0) = 0 \text{ for } x\in (0, L]
\end{align*}
and
\begin{align*}
  \rho(0,t) = \rho_{\rm{max}} \ , \; \left.\pd{\rho(x,t)}{x}\right|_{x=L}=0.
\end{align*}

The solution can then be found using standard methods (App.~\ref{sec:sourceAsBoundaryApp}) and expressed as a series:
\begin{equation}\label{eq:sourceasboundary}
    \frac{\rho(x,t)}{\rho_{\rm{max}}} =1- \frac{4}{\pi}\sum_{n=1}^{\infty}\frac{\sin\left(\frac{(2n-1)}{2}\pi \frac{x}{L}\right)}{2n-1}e^{-\left(\frac{(2n-1)\pi}{2L}\right)^2D\;t} \ .
\end{equation}

\begin{figure}
    \includegraphics[width=0.48\textwidth]{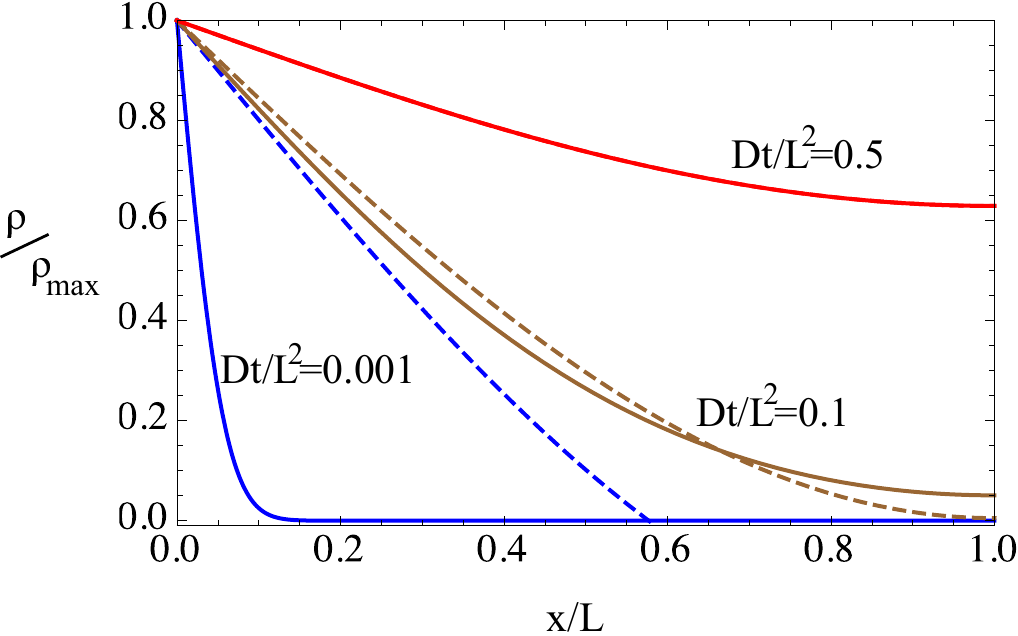}
    \caption{Analytic food distribution  for times $t=0.001 L^2/D$ (blue line, bottom), $t=0.1 L^2/D$ (brown line, middle), $t=0.5 L^2/D$ (red line, top). The solid lines are the full solution (\eqref{sourceasboundary}) and the dashed lines are the approximation (\eqref{sourceasboundary-approx}). The dashed line for $t=0.5 L^2/D$ overlaps with the solid line and is not visible. }
    \label{fig:rhoWithFoodAsSource-analytically}
\end{figure}

Keeping only the dominant first term ($n=1$), the solution reads:
\begin{equation}
    \rho(x,t) \approx \rho_{\rm{max}}\left[1-\frac{4}{\pi}\sin\left(\frac{\pi x}{2L}\right)e^{-\left(\frac{\pi}{2L}\right)^{2}D\;t}\right] \ .
    \label{eq:sourceasboundary-approx}
\end{equation}
In Fig.~\ref{fig:rhoWithFoodAsSource-analytically}, we plot \eqref{sourceasboundary} (solid line) and \eqref{sourceasboundary-approx} (dashed line) for times $t=0.001 L^2/D$, $t=0.1 L^2/D$ and $t=0.5 L^2/D$.
We see that for large enough times, \eqref{sourceasboundary-approx} is an excellent approximation of \eqref{sourceasboundary}. Note also that the overall shape of the spatial distribution of the food and its time dependence qualitatively match the small $\lambda$ simulations displayed in \figref{food_spatial_distribution_1}.
However, this continuum approach cannot capture the very interesting individual ant food distribution dynamics shown in the histograms of \figref{food_spatial_distribution_1}. For example, we see that in several regimes the mean food at a given location is not a good indicator of the actual food concentration of individual ants at that location (cf. \figref{food_spatial_distribution_1}(a${}_1$)). We see that although $\langle c(t=10)\rangle_{x=0.1} \approx 0.75$, there are no individual ants in the bin $x \in [0.05, 0.1]$ with $c_i(t=10) = 0.75$. Instead, at that location, the food is distributed bimodally, with approximately 75\% of the ants at capacity and 25\% empty.

From \eqref{sourceasboundary}, we can calculate $\langle c(t) \rangle$ by integration,
keeping only the slowest decaying term. The approximate average food in the nest is then found to be:
\begin{equation}
\langle c(t) \rangle \approx  c_{\rm{max}} \left(1-\frac{8}{\pi ^2}     e^{- \left(\frac{\pi}{2L}\right)^2  D\; t    } \right)
\end{equation}
and the mean squared distance:
\begin{equation}
\overline{\rm{MSD}}(t) \approx 1-\frac{96(\pi-2) }{\pi ^4} e^{- \left(\frac{\pi}{2L}\right)^2 D\; t}.
\end{equation}

\begin{figure}
     \includegraphics[width=0.48\textwidth]{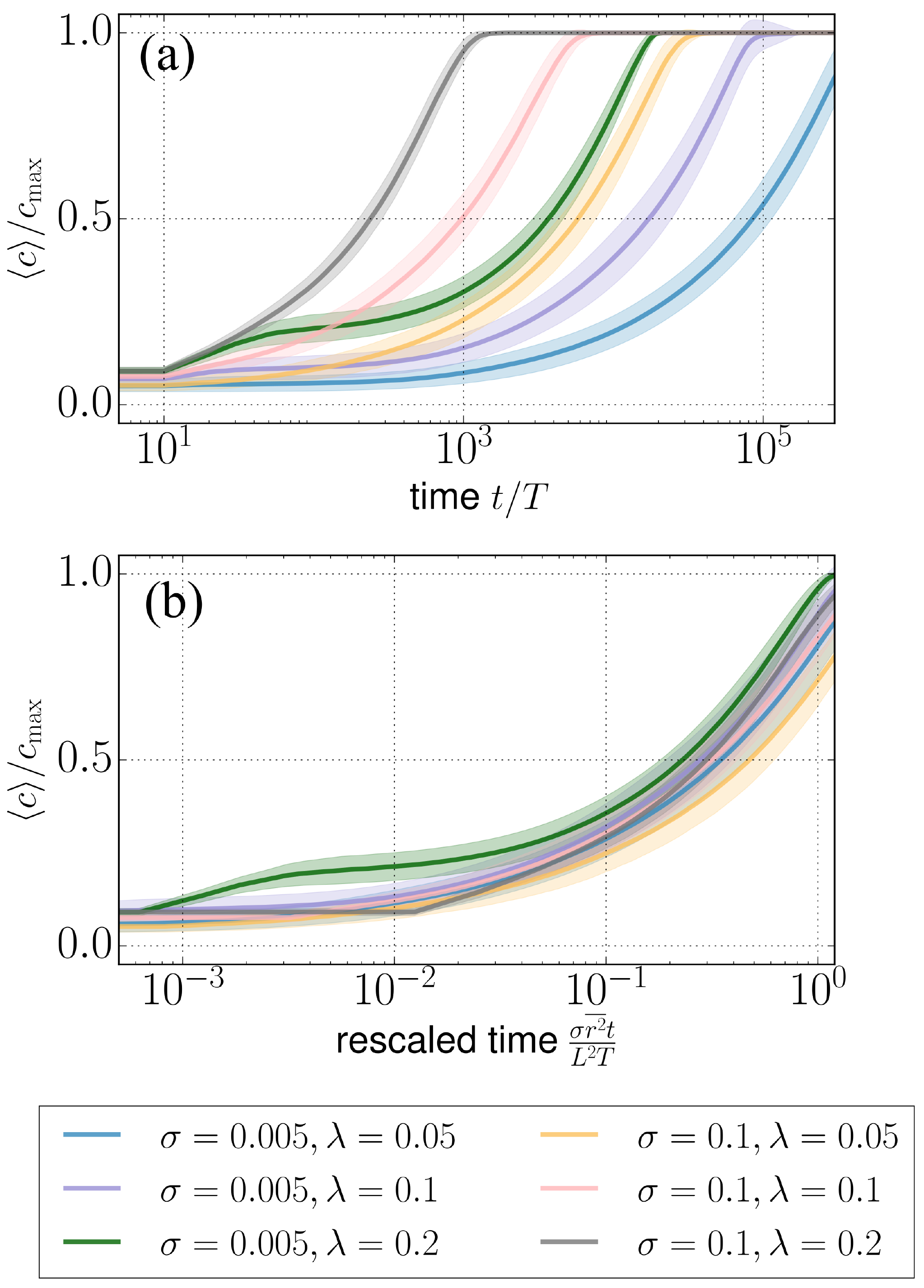}
     \caption{Average food concentration $\langle c \rangle/c_{\rm{max}}$ (a) as a function of  time $t/T$ (b) and as a function of rescaled time $\frac{\sigma\overline{r^2}t}{L^2T}$ for various values of $\sigma$ and $\lambda$. After rescaling the time, the curves collapse on the same master curve. The width of the shaded area around each line indicates the standard deviation. 
     }
     \label{fig:mean_food_vs_diffusion_time}
\end{figure}
\begin{figure}
     \includegraphics[width=0.48\textwidth]{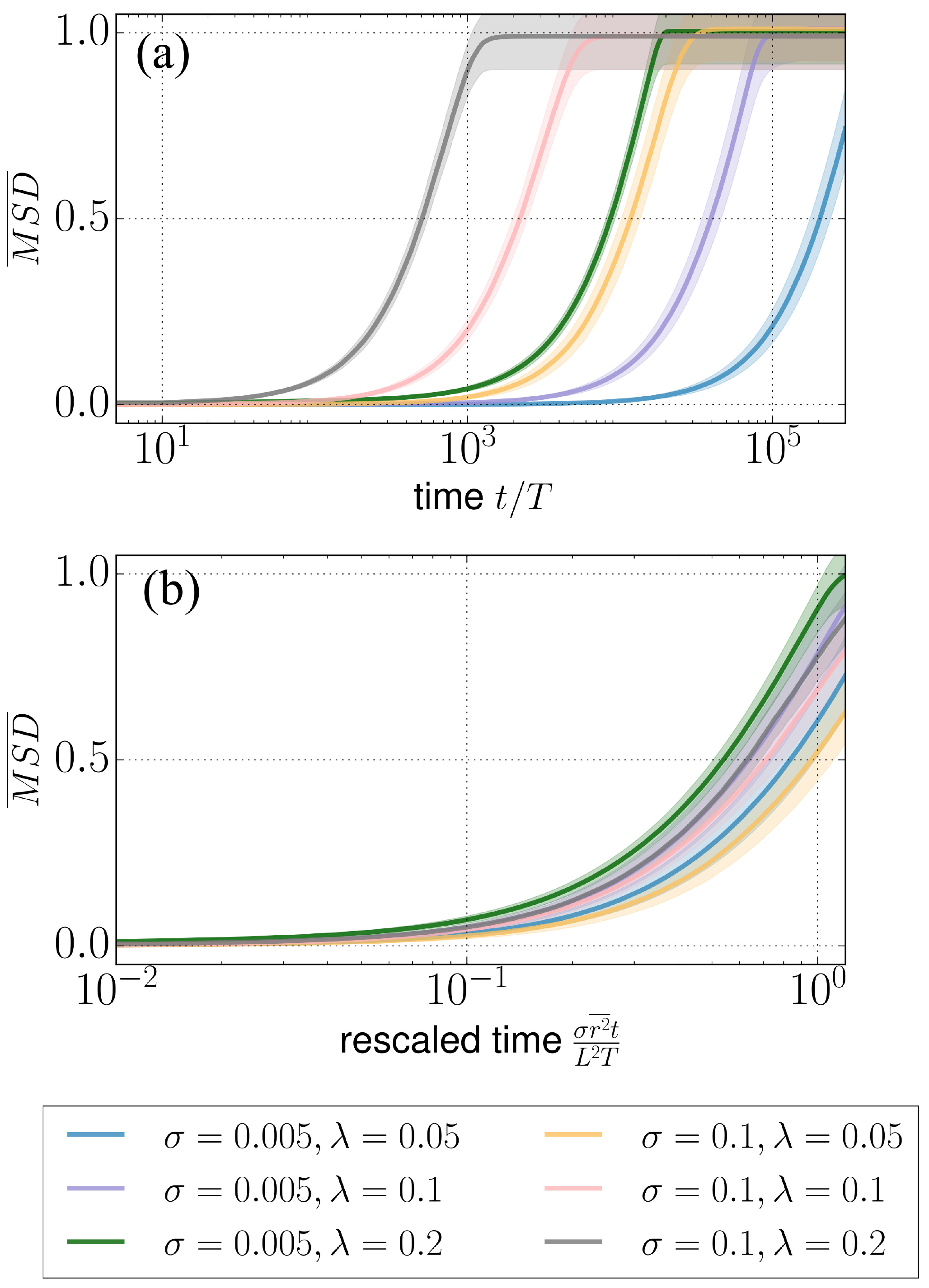}
    \caption{Mean squared food distance $\overline{\text{MSD}}$ (a) as a function of time $t/T$ (b) and as a function of rescaled time $\frac{\sigma\overline{r^2}t}{L^2T}$ for various values of $\sigma$ and $\lambda$. After rescaling the time, the curves collapse on the same master curve. The width of the shaded area around each line indicates the standard deviation.
    }
    \label{fig:msd_vs_diffusion_time}
\end{figure}

Associating the resulting exponential decay parameter $\left(\frac{\pi}{2L}\right)^{2}D$ with the decay parameter $\gamma$ of the exponential source term in 
\eqref{food-decay-after-transition} yields
\begin{align}\label{eq:scalingD}
    \gamma = \left(\frac{\pi}{2L}\right)^{2}D \ .
\end{align}

Using \eqref{diffusionConstant} we obtain
\begin{align}\label{eq:scaling}
    \gamma \sim \frac{\sigma \overline{r^2}}{T L^2} \ .
\end{align}
For equidistant agents with nearest neighbor interactions, using $\overline{r^2}=(L/N)^2$, this reduces to 
\begin{align}
    \gamma \sim \frac{\sigma}{T N^2} \ .
    \label{eq:timescale}
\end{align}

Thus, \eqref{scaling} provides the characteristic time scale for the trophallactic process in the nest.  In \figref{mean_food_vs_diffusion_time} and \figref{msd_vs_diffusion_time} we plot $\langle c \rangle$ and  $\overline{\rm{MSD}}(t)$ respectively for $\sigma=0.005$ and $0.1$; and $\lambda=0.05,\; 0.1$ and $0.2$. In panel (a) we plot the quantities as a function of $t/T$, whereas in panel (b), we rescale time using $\gamma$ in \eqref{scaling}. We find a relatively good collapse of the data after an initial period where the dynamics is dominated by direct source food uptakes from agents in range of the source. This indicates that despite the oversimplifications of the analytic model, the timescale $L^2T/\sigma \overline{r^2}$ captures the dynamics of the discrete, agent-based model.

\subsection{The food source as a source term.}
In this section, we consider the alternative case of modeling the food source as a source term of the diffusion equation. For computational ease, in this section we consider a 1d semi-infinite nest, with the source being located at $x=0^+$. The diffusion equation that governs the dynamics of the system reads:
\begin{dmath}
    \label{eq:diffusion}
    \pd{}{t}\rho(x,t) = D \pd[2]{}{x}\rho(x,t) + q(x,t),
\end{dmath}
where $\rho(x,t)$ is the food density, $D$ the diffusivity, and $q(x,t)$ the source term. 

The food is taken up at the single point $x=0$ at an exponential rate with unknown exponent $\gamma$, similarly to what is observed in the experiments~\cite{Greenwald2015}. 
Again, we focus on the initial stages of the trophallactic process, assuming a constant $D$ after \eqref{diffusionConstant}, and that ants have a finite capacity. We estimate a bound for the exponent $\gamma$.
As we showed in Section~\ref{sec:LargeR}, the amount of food remaining at the nest entrance decays approximately exponentially in time (\eqref{mean-field-food-mean}). If the total amount of food available initially is  $f_{\rm{max}}$ (as in \eqref{rho-max}), then the total amount of food still available at the source at time $t$ is $f_{\rm{max}} e^{-\gamma t} $ and the source term reads
\begin{align}
    \label{eq:source}
    q(x,t) &= - \pd{}{t} f_{\rm{max}} e^{-\gamma t} 2 \delta(x) \nonumber \\
           &= 2 \gamma f_{\rm{max}} e^{-\gamma t} \delta(x) \ .
\end{align}
where $\delta(x)$ the Dirac delta distribution, as the nest entrance is located at $x=0^+$. For normalization purposes we include a factor of $2$ in \eqref{source} and assume $\int_0^{\infty} \delta(x) dx = 1$.

\eqref{diffusion} can be non-dimensionalized by choosing the time scale $s_{t}$, length scale $s_{x}$ and food scale $s_{\rho}$ to be
\begin{align*}
    s_{t} = \frac{1}{\gamma}, \quad s_{x} = \sqrt{\frac{D}{\gamma}} \quad \text{and} \quad s_{\rho} =2 f_{\rm{max}}/s_{x}  \ .
\end{align*}
Denoting a non-dimensional variable as $\tilde{X} \coloneqq \frac{X}{s_{X}}$, \eqref{diffusion} becomes
\begin{dmath}
    \label{eq:diffusion-nondimensional}
    \pd{}{\tilde{t}}\tilde{\rho}(\tilde{x},\tilde{t}) = \pd[2]{}{\tilde{x}}\tilde{\rho}(\tilde{x},\tilde{t}) + e^{-\tilde{t}} \delta(\tilde{x}) \ .
\end{dmath}
With the boundary and initial conditions
\begin{align*}
    \rho(x,0) = 0 \text{ for } x \ge 0
\end{align*}
and
\begin{align*}
 	\left.\pd{\rho(x,t)}{x}\right|_{x=0}=0,
\end{align*}
the general solution of the semi-infinite system in non-dimensional terms is
\begin{dmath}
    \label{eq:general-solution-nondimensional}
    \tilde{\rho}(\tilde{x},\tilde{t}) = \int_{0}^{\infty} \dif \tilde{x}' \int_{0}^{\tilde{t}} \dif \tilde{t}' \tilde{K}(\tilde{x},\tilde{x}',\tilde{t}-\tilde{t}')e^{-\tilde{t'}} \delta(\tilde{x'})
\end{dmath},
where 
\begin{dmath*}
    \tilde{K}(\tilde{x},\tilde{x}',\tilde{t}) \coloneqq \frac{1}{\sqrt{4\pi \tilde{t}}}e^{-\frac{(\tilde{x}-\tilde{x}')^{2}}{4 \tilde{t}}}
\end{dmath*}
is the heat kernel of \eqref{diffusion-nondimensional}.
Dropping the tildes, the non-dimensional solution finally becomes
\begin{dmath}\label{eq:fullRho}
    \rho(x,t) = e^{-t} \left[- \frac{1}{2}\Ima\left(e^{ix} \erfc\left(\frac{x}{2\sqrt{t}}+i\sqrt{t}\right)\right) \right],
\end{dmath}
where $\erfc(x)=1 - \erf(x)$ is the complementary error function, and $\rm{Im}(x)$ denotes the imaginary part of $x$ (see App.~\ref{sec:sourceAsBoundaryApp} for derivation).
Some simpler analytic approximations can be derived by considering the limits of the complementary error function.
For large times and far from the source,
\begin{align*}
    \rho(x,t) \approx \frac{1}{2} \sqrt{\frac{t}{\pi}} \frac{1}{\left(\frac{x^2}{4 t}+t\right)}e^{-\frac{x^{2}}{4t}}
\end{align*}
approximates \eqref{fullRho} well, whereas for short times and close to the source 
\begin{align*}
    \rho(x,t) \approx 2 \sqrt{\frac{t}{\pi}} e^{-\frac{x^{2}}{4t}} -\left| x\right|  e^{-t}
\end{align*}
is a good approximation.

We now explore the relationship of $\gamma$ to $D$. The higher $\gamma$ is, the faster the nest absorbs food from the source. However,  
a very large $\gamma$ would lead to a very large density $\rho(x,t)$ in the vicinity of the nest entrance, which contradicts the finite carrying capacity. 
In the rest of this section, we derive an upper limit for $\gamma$ that is consistent with the finite carrying capacity of the system $\rho_{{\rm max}}$.


Going back to dimensional notation, \eqref{fullRho} evaluated near the source ($x \rightarrow 0$) reads:
\begin{align}
    \label{eq:solution-at-source}
    \rho(0,t) = - f_{\rm{max}}\sqrt{\frac{\gamma}{D}}e^{-\gamma t}\Ima\left( \erfc\left(i\sqrt{\gamma t}\right)\right) \ .
\end{align}
Since $\gamma t \geq 0$,
\begin{align*}
    \Ima\left( \erfc\left(i\sqrt{\gamma t}\right)\right) = -\frac{2}{\sqrt{\pi}}\int_0^{\sqrt{\gamma t}}e^{z^2}\dif z \ ,
\end{align*}
and \eqref{solution-at-source} becomes
\begin{align*}
    \rho(0,t) = \frac{2 f_{\rm{max}}}{\sqrt{\pi}}\sqrt{\frac{\gamma}{D}}e^{-\gamma t}\int_0^{\sqrt{\gamma t}}e^{z^2}\dif z \ .
\end{align*}
The finite crop capacity is reached at a time $t_{f}$, such that
\begin{align*}
    \rho(0,t_{f}) = \rho_{\rm{max}}.
\end{align*}
The time $t_f$ when this equality is fulfilled depends on $\gamma$ and $\sigma$. 
A large $\gamma$ or a small $\sigma$ will lead to fast food saturation near the source.

In fact, for every time $t$, the food density near the source has to be less than or equal to $\rho_{\rm{max}}$,
\begin{align}
    \nonumber
    \rho(0,t) &\le \rho_{\rm{max}} \\
    \nonumber
    \Leftrightarrow \tilde{\rho}(0,\tilde{t}) &\le \frac{\rho_{\rm{max}}}{s_{\rho}} = \frac{1}{2 L}\sqrt{\frac{D}{\gamma }}\\
    \nonumber
    \Leftrightarrow \gamma &\le  \frac{1}{\tilde{\rho}(0,\tilde{t})^2 } \frac{D}{ (2 L)^2}\\
\label{eq:rho-tf-equal-rho-max}
\end{align}
where the tildes denote dimensionless quantities.

By numerically evaluating the dimensionless density at the origin (e.g. using Mathematica), we find that $\tilde{\rho}(0,t)$ initially increases, reaches the maximum value of $\tilde{\rho}(0,t) \simeq 0.31$ and then decreases as the finite amount of food at the source diffuses to infinity. Every value of $\tilde{\rho}(0,t)$ results in a different upper limit for $\gamma$ through \eqref{rho-tf-equal-rho-max}. Hence, if $\gamma$ is constant throughout the process, the upper bound for $\gamma$ should be equal to $\min_{\tilde t} (\frac{1}{\tilde{\rho}(0,\tilde{t})^2 } \frac{D}{ (2 L)^2})$, or
\begin{align*}
   \gamma &\lesssim 10.4 \frac{D}{(2L)^2} \ .
\end{align*}
This is consistent with \eqref{scalingD} derived for the finite system. The $D/L^2$ dependency of $\gamma$ is not surprising as it could have been easily predicted by considering the dimensionless groups that can be constructed with the equation parameters, $\gamma$, $D$ and $L$. However, note that $L$ is not an explicit length scale of the semi-infinite system presented in this section, but comes in 
through the equation relating the total amount of food $f_{\rm max}$ and the saturation density $\rho_{\rm{max}}$ and can be re-expressed as $ \gamma \lesssim 10.4 \frac{D \rho_{\rm{max}}^2}{(2 f_{\rm{max}})^2}$. 

Thus, we have shown how the diffusion coefficient and the implicit system length scale provide an upper bound for the overall food uptake by the nest.

\section{Half time and variance}\label{sec:Results}

\begin{figure} 
    \centering
    \includegraphics[width=0.45\textwidth]{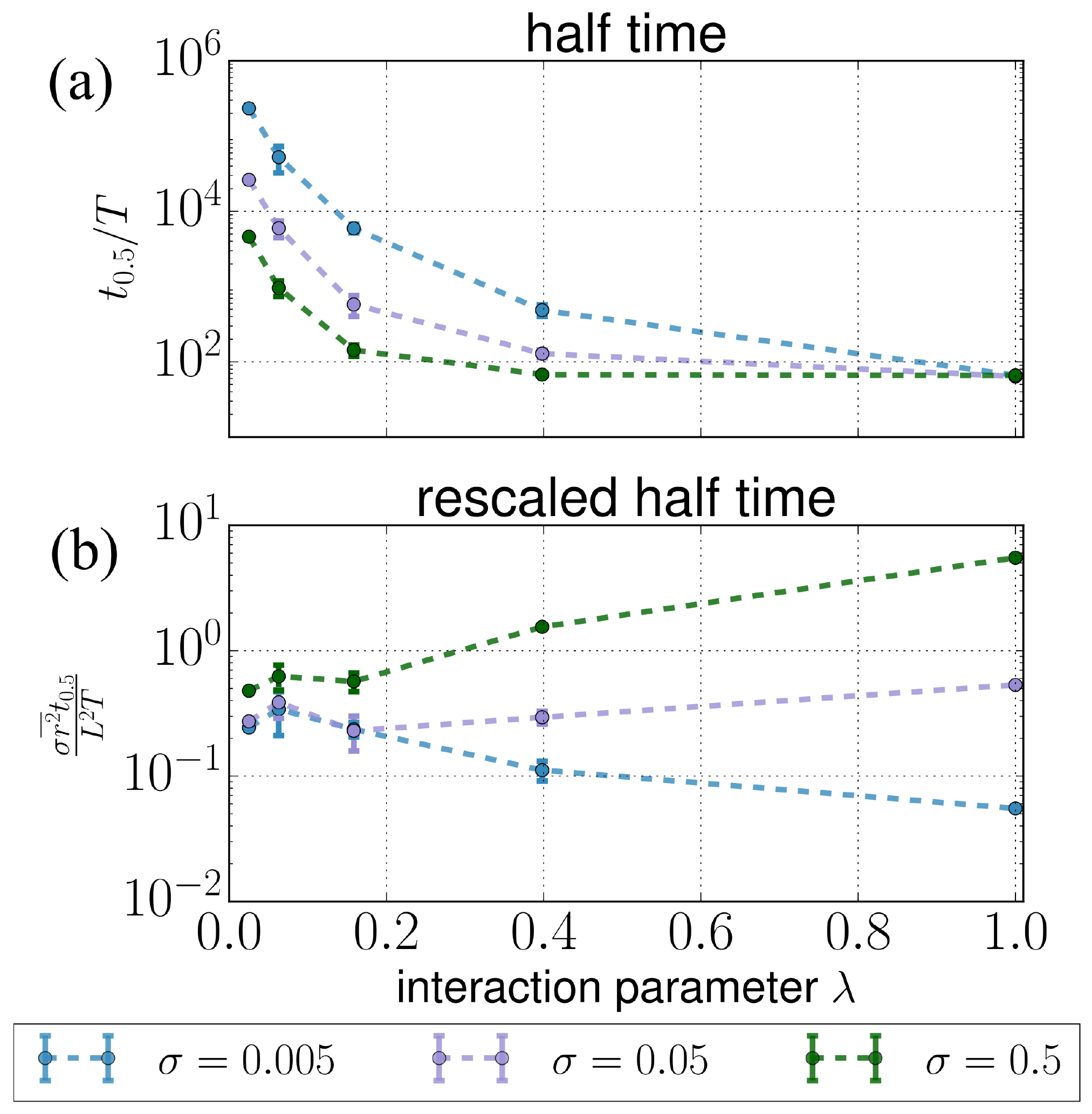} \\
    \caption{ \label{fig:50full} The time $t_{0.5}$ until
    the average food in the colony reaches $0.5 c_{\mathrm{max}}$.
(a) Half time $t_{\rm{0.5}}$ as a function of $\lambda$ for $\sigma=0.005$ (blue, top curve), $\sigma=0.05$ (purple, middle curve) and $\sigma=0.5$ (green, bottom curve). (b) Rescaled half time $\frac{\sigma \overline{r^2} t_{\rm{0.5}}}{T L^2}$ as a function of $\lambda$ for $\sigma=0.005$ (blue, bottom curve), $\sigma=0.05$ (purple, middle curve) and $\sigma=0.5$ (green, top curve). The half time decreases with increasing $\lambda$ and $\sigma$. The $\sigma$ and $\lambda$ dependence of  $t_{\rm{0.5}}$ can be for the most part explained by the scaling of \eqref{scaling}. For high $\lambda$, the $\sigma$ dependence gradually disappears, and the scaling of \eqref{scaling} fails to explain the behavior. In that limit, the relevant timescale is that of \eqref{mean-field-food-mean}.  }
 \end{figure}

In \figref{50full} we plot the time $t_{\rm{0.5}}$ it takes for the nest to acquire half of the available food. We see that the time decreases with increasing $\sigma$, the proportion of food the ants attempt to exchange at each interaction, and $\lambda$, the normalized interaction radius that is a proxy of how well mixed the colony is (or alternatively how broad the fidelity zones of the ants). We see that the half-time $t_{\rm{0.5}}$ dependence on $\sigma$  declines with increasing $\lambda$. 

As shown in \figref{50full}, the half time is approximately  $t_{\rm{0.5}} \approx 0.4 \frac{T L^2}{\sigma \overline{r^2}}$ for small interaction radii.
For high interaction radii, the half time scales as $t_{\rm{0.5}} \sim NT$, as can be deduced from \eqref{mean-field-food-mean}.

\begin{figure}
    \centering
    \includegraphics[width=0.45\textwidth]{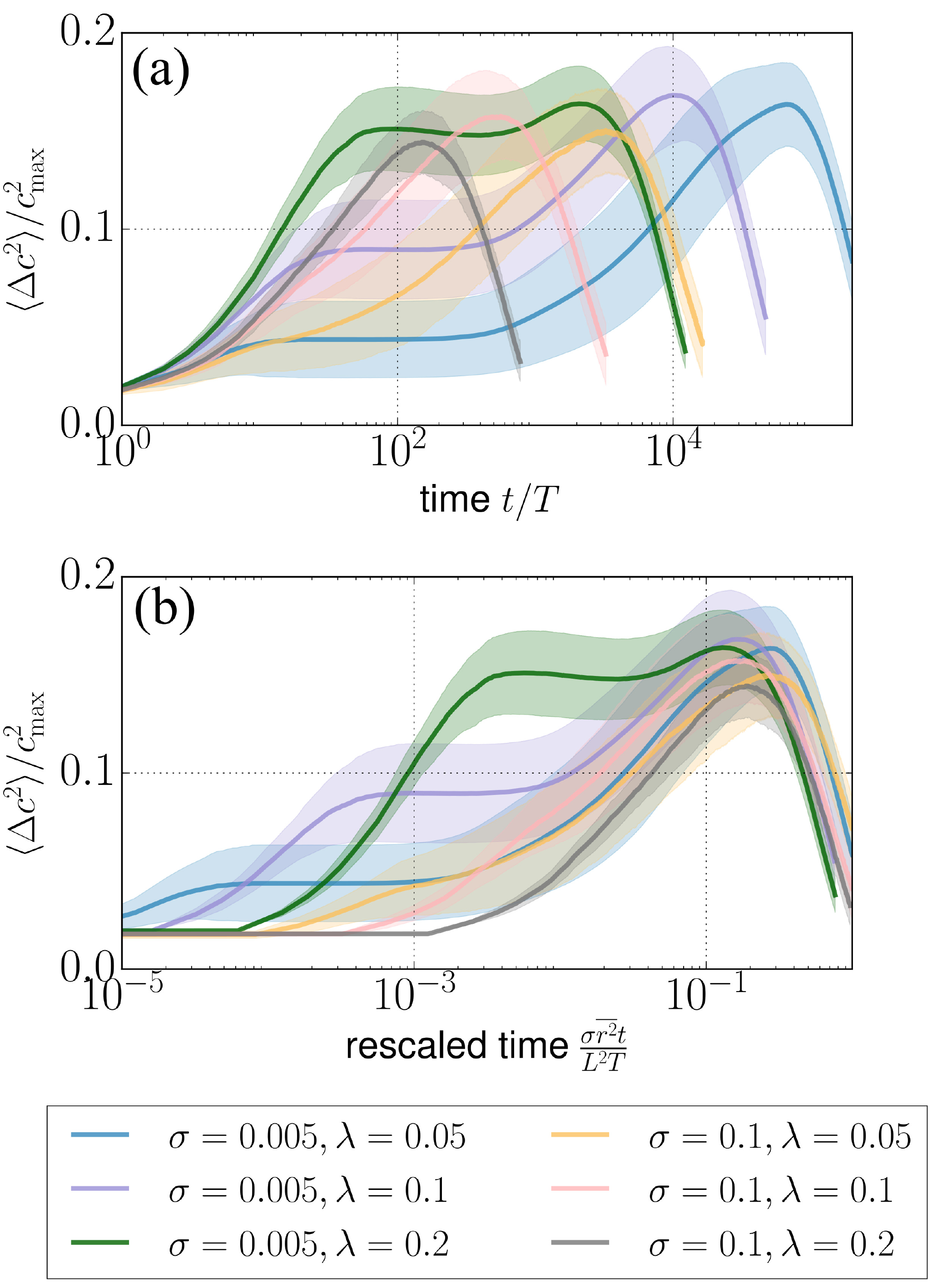} \\
    \caption{\label{fig:variance} Normalized food variance $\langle \Delta c^2 \rangle / c_{\rm{max}}^2$ (a) as a function of time $t/T$ and (b) as a function of rescaled time $\frac{\sigma\overline{r^2}t}{L^2T}$ for various values of $\sigma$ and $\lambda$. After rescaling the time, the curves collapse on the same master curve at the later stages of the dynamics. The variable shapes and plateau sizes of the initial stages of the dynamics reflect the complicated spatial food distributions along the length of the nest, presented in Fig.~\ref{fig:food_spatial_distribution_1}. }
 \end{figure}

The food variance $\langle \Delta c^2 \rangle$ in the colony is a proxy of how well the available food is distributed among the agents in the nest. It is displayed in \figref{variance}. After an initial increase, the variance plateaus for some time, until it reaches a maximum approximately at a time $0.3 \frac{T L^2}{\sigma \overline{r^2}}$,  roughly when the colony is becoming half full. After that time, the variance is monotonically decreasing, as the food is distributed better, due to the emerging saturation of the colony. 

At the beginning of the food dissemination process, the behavior of the variance, showcased in Fig.~\ref{fig:variance}, reflects the food uptake dynamics when there is no saturation near the source. During the initial sharp increase of the small $\sigma$ curves, the ants inside the interaction radius of the source, initially all empty,  quickly take up food at a rate similar to the mean-field model. As $\sigma$ is small, the rate at which the food diffuses away from that zone is slower than the fast uptake of food at the entrance. Consequently,  $\langle \Delta c^2 \rangle$ increases quickly, and a growing proportion of the ants near the entrance reaches saturation. The food uptake rate of the area near the source then becomes approximately equal to the rate at which the food diffuses out of that zone, and the slope of $\langle \Delta c^2 \rangle$ decreases. Eventually, when that area is close to saturation, it acts as an effective boundary condition $\rho(R,t)=\rho_{\rm{max}}$, and the food diffuses to the rest of the nest as a propagating front (see e.g. Fig.~\ref{fig:food_spatial_distribution_1} ($\rm{a}_1$)-($\rm{a}_3$)). This clear separation of the two timescales happens only for small $\sigma$. For large $\sigma$, the plateau disappears.

Since for relatively small $\sigma$ the front between the ants that are at capacity (to the left of the propagating front of Fig.~\ref{fig:food_spatial_distribution_1}) is relatively sharp, we can approximate the normalized variance as $\langle \Delta c^2 \rangle / c_{\rm{max}^2} \approx n/N - (n/N)^2$, where $n$ is the number of ants at capacity. This is maximized when $n=N/2$, i.e. when the colony is half full, in agreement with the simulation results of \figref{variance} and \figref{mean_food_vs_diffusion_time}.

\section{Discussion}\label{sec:discussion}
We have presented a simple agent-based model of trophallaxis, the mouth to mouth dissemination of food in an
ant colony. The agents (ants) in our model have finite capacity and can move in a finite nest. Their interaction is confined to a finite zone (spatial fidelity zone), which, depending on the parameter selection, can extend to the whole nest. The agents pick an interaction partner at random and donate a fixed percentage of their crop content, limited by the recipient's crop capacity. Agents that can reach the entrance of the nest are picked at random to fill their crop at capacity with food from the source. Our model describes the trophallactic process in the simplest possible terms, assuming
stochastic effects over planning or strategy wherever possible. 

We then proceed to consider analytically tractable simplifications of the agent-based model
in order to understand the model behavior for a well mixed colony (i.e. quickly moving agents or large interaction radius), and for nests with spatial fidelity zones. For both cases, we have derived characteristic time scales for the food uptake rate that allow us
to describe the collective trophallactic dynamics. The time scales generally depend on parameters such as number of ants, the typical time between interactions, and the fraction of food transmitted at each trophallactic event. We find that for localized ants with spatial fidelity zones, the characteristic food uptake timescale decreases as the percentages of food exchanged or the spatial fidelity zones become larger. This acceleration of food uptake is also reflected in the timescales of how the food is spread through the nest and distributed among the colony. We also find that this timescale is more generally an upper limit of how fast a trophallactic system of agents with finite capacity can absorb food. This leads us to conclude that, in the case of spatial fidelity zones of limited extent, a drastically different strategy would be needed to significantly accelerate food uptake, e.g. by having dedicated foragers that enter the nest and freely move to all locations to deliver food rather than staying close to the entrance and relying on diffusion for food dissemination. 

For increasing spatial fidelity zones, the amount of food being exchanged plays a progressively smaller role in determining the characteristic timescales. 

Finally, for a well mixed colony, where the interactions between the ants happen at random, there is minimal influence of the proportion of food exchanged between ants on how fast the colony can absorb food. The food uptake is then primarily determined by how frequently the source is visited.

All of the free parameters that we consider (number of ants, system size, proportion of food exchanged, extent of spatial fidelity zone and interaction duration time) are easy to measure and to compare to experiments such as~\cite{Greenwald2015}. With our work, we have laid the theoretical groundwork for a quantitative description of trophallaxis in ants and other eusocial insects in terms of simple observables. 

The behavior of ant colonies in a realistic setting departs in many ways from our basic stochastic model. Real ant colonies can exhibit division of labor, information exchange introducing non-random behavior, or non-trivial trophallaxis strategies. However, our model and predictions can provide a useful benchmark to assess to what level the observed food uptake rates and efficiency in food distribution is due to stochastic effects or ingeniously applied trophallactic strategies by the ant colony.

\begin{acknowledgments}
EK acknowledges support from the Burroughs Wellcome Fund and the NSF award PHY-1554887. She also wishes to thank J. Walker for useful discussions.
\end{acknowledgments}

\appendix

\section{Solving the diffusive limit, source as a boundary condition}\label{sec:sourceAsBoundaryApp}
The system solved in Sec.~\ref{sec:source-as-boundary} is equivalent to:
\begin{align*}
    x \in [0,2L]  \;  \text{for}  \; \rho(x,0)=0
\end{align*}
and
\begin{align*}
  \rho(0,t) = \rho_{\rm{max}},  \qquad \rho(2L,t)=\rho_{\rm{max}}.
\end{align*}
Transforming to $\rho' = \rho_{\rm{max}} - \rho$ turns the boundary and initial conditions into
\begin{align*}
    \rho'(0,t) = 0 \ , \quad \rho'(2L,t)=0 \quad \text{and} \quad \rho'(x,0)=\rho_{\rm{max}} \ .
\end{align*}
Solving the diffusion equation with these boundary conditions is known as the problem of cooling of a rod and can be done through separation of variables. The solution reads
\begin{align*}
    \rho'(x,t) = \frac{4\rho_{\rm{max}}}{\pi}\sum_{n=1}^{\infty}\frac{\sin\left(\frac{(2n-1)}{2}\pi \frac{x}{L}\right)}{2n-1}e^{-\left(\frac{(2n-1)\pi}{2L}\right)^2Dt} \ .
\end{align*}

We transform back to $\rho = \rho_{\rm{max}} - \rho'$ to get \eqref{sourceasboundary}.

A simple integration can give us the average food
\begin{equation*}
\langle c(t) \rangle = c_{\rm{max}} \left(1-\frac{8}{\pi ^2}   \sum _{n=1}^{\infty}  \frac{1}{(2 n-1)^2}   e^{ -(2 n-1)^2  \frac{\pi ^2}{4L^2}  D t    } \right)
\end{equation*}
and the mean squared distance:
\begin{equation*}
\overline{\rm{MSD}}(t) = 1+\frac{96 }{\pi ^4}\sum _{n=1}^{\infty} \frac{2+ \pi  (-1)^n (2 n-1) }{(2n-1)^4}e^{ -(2 n-1)^2 \frac{\pi ^2}{4L^2} D t}.
\end{equation*}

\section{Solving the diffusive limit, explicit source term}
For completeness, here we outline the  derivation of \eqref{fullRho}. In this derivation, we assume that the system is infinite (as opposed to semi-infinite, as in the main text), and that the source is at $x=0$.
The solution for the semi-infinite case can be easily obtained from the full system by considering only $x>0$ and adjusting the normalization. 

Dropping the tildes, the solution of \eqref{diffusion-nondimensional} for $x \in (-\infty , \infty)$ reads:
\begin{dmath*}
    \rho(x,t) = \frac{1}{\sqrt{4\pi}} \int_{-\infty}^{\infty} \dif x' \int_{0}^{t} \dif t' \frac{e^{-\frac{(x-x')^{2}}{4 (t-t')}- t'}}{\sqrt{t-t'}}\delta(x') \\
    = \frac{1}{\sqrt{4\pi}} \int_{0}^{t}  \frac{e^{-\frac{x^{2}}{4 (t-t')}- t'}}{\sqrt{t-t'}} \dif t' \\
    = \frac{1}{\sqrt{4\pi}} \left[-\frac{\sqrt{\pi}}{2i}e^{-t}\left(e^{-ix}\left(1+\erf \left(-\frac{x}{2\sqrt{t-t'}}+i\sqrt{t-t'}\right)\right) \\ + e^{ix}\left(-1+\erf \left(\frac{x}{2\sqrt{t-t'}}+i\sqrt{t-t'}\right) \right)\right)\right]_{0}^{t}
\end{dmath*}.
The upper boundary ($t'=t$) of the time integral evaluates to zero for $x \geq 0$. For $x<0$, it evaluates to 
\begin{align*}
    \sqrt{4\pi}e^{-t}\sin(x) \ .
\end{align*}
The lower boundary ($t'=0$) evaluates to
\begin{align*}
    &-\frac{\sqrt{\pi}}{2i}e^{-t} \Bigg(e^{-ix}\left(1+\erf \left(-\frac{x}{2\sqrt{t}}+i\sqrt{t}\right)\right) \\
    &\qquad + e^{ix}\left(-1+\erf \left(\frac{x}{2\sqrt{t}}+i\sqrt{t}\right) \right)\Bigg) \ .
\end{align*}
Through making use of $\erfc(z) \coloneqq 1-\erf(z)$ and denoting the complex conjugate of $z$ with $\bar{z}$, the solution can be concisely written as:
\begin{dmath*}
    \rho(x,t) = e^{-t} \left[\sin(x) \Theta(-x) + \frac{1}{4i}\left( \conj{u} \erfc(\conj{w}) - u \erfc(w)\right) \right]
\end{dmath*},
where
\begin{dseries}
\begin{math}
    u \coloneqq e^{ix}
\end{math}
and
\begin{math}
    w \coloneqq \frac{x}{2\sqrt{t}}+i\sqrt{t}
\end{math}.
\end{dseries}
Since $\conj{u} \erfc(\conj{w}) - u \erfc(w) = -2i \Ima(u \erfc(w))$, the solution finally becomes
\begin{dmath}\label{eq:fullSolutionrho}
    \rho(x,t) = e^{-t} \left[\sin(x) \Theta(-x) - \frac{1}{2}\Ima\left(e^{ix} \erfc\left(\frac{x}{2\sqrt{t}}+i\sqrt{t}\right)\right) \right] \ .
\end{dmath}
The solution in the main text (\eqref{fullRho}) can be obtained from this solution by dropping the $\Theta(x)$ term. Note that \eqref{fullSolutionrho}, integrated over $x \in (-\infty, \infty)$ at $t \to \infty$ will eventually yield $\int_{-\infty}^{\infty} \rho(x,t \to \infty)=2f_{\rm max}$ in dimensional terms.

\section{Averaged interaction radius}
\label{sect:interaction-radius}

The derivation of the ensemble averaged squared distance between pairs of interacting ants $\overline{r^{2}}(R)$ can be split into two parts, depending on the size of the interaction radius $R$:
\begin{enumerate}
    \item The ant's interaction range overlaps only with one system boundary or does not overlap with the boundaries: $R \in [0,\frac{L}{2})$.
    \item The ant's interaction range either overlaps with one or with both system boundaries: $R \in [\frac{L}{2},L]$.
\end{enumerate}

\subparagraph{1. Case $\bm{R \in \left[0,\frac{L}{2}\right)}$:}
The average squared distance $\overline{r_{f}^{2}}$ of a point $r$ in the interaction range $[-R,R] \subset \mathbb{R}$ (e.g. one ant) to the center of the interval $r=0$ (e.g. the other ant) free of boundaries can be calculated as the mean of the uniform distribution $u(I)$ of points in the interval $I = [-R,R]$. The uniform distribution is
\begin{align*}
    u(I) \coloneqq \frac{1}{\int_{I} \dif r} = \frac{1}{\abs{I}} = \frac{1}{2R} \ ,
\end{align*}
where $\abs{I}=2R$ denotes the length of the interval. \\
Using this, the boundary free average squared distance can be calculated as
\begin{align}
\label{eq:r-squared-f}
    \overline{r_{f}^{2}} = \int_{I} r^{2} u(I)\dif r = \frac{\int_{-R}^{R} r^{2} \dif r}{\int_{-R}^{R} \dif r} = \frac{1}{2R} \int_{-R}^{R} r^{2} \dif r = \frac{R^{2}}{3}\, .
\end{align}
The same approach can be used to calculate the average squared distance $\overline{{r}_{b}^{2}}(s)$ of a point $r$ in the truncated interaction range $[-s,R] \subset [-R,R]$ (e.g. one ant) to the original center of the interval at $0$  (e.g. the other ant), when the interaction range overlaps with one system boundary. Without loss of generality, this boundary is put to the left in the calculation. Due to symmetry, the result remains the same for an overlap with the right system boundary.
\begin{align}
\label{eq:r-squared-b}
    \overline{{r}_{b}^{2}}(s) = \frac{1}{s+R} \int_{-s}^{R} r^{2} \dif r = \frac{R^{3}+s^{3}}{3(s+R)}\, .
\end{align}
In order to calculate the system wide average over these average squared distances, the system interval $[0,L]$ needs to be split up into the following three regions:
\begin{enumerate}
    \item The average squared distances resulting from interaction ranges overlapping with the left system boundary: $\frac{1}{R}\int_{0}^{R}\overline{{r}_{b}^2}(x) \dif x\, $.
    \item The average squared distances resulting from non overlapping interaction ranges in the central region: $\frac{1}{L-2R}\int_{R}^{L-R}\overline{{r}_{f}^2} \dif x\, $.
    \item The average squared distances resulting from interaction ranges overlapping with the right system boundary: $\frac{1}{R}\int_{L-R}^{L}\overline{{r}_{b}^2}(L-x) \dif x\, $.
\end{enumerate}
These three regions can then be averaged using their lengths as weights to obtain the system wide average:
\begin{dmath*}
    \overline{r^{2}} = \frac{1}{L} \left(\int_{0}^{R}\overline{{r}_{b}^2}(x) \dif x + \int_{R}^{L-R}\overline{{r}_{f}^2} \dif x + \int_{L-R}^{L}\overline{{r}_{b}^2}(L-x) \dif x \right)
\end{dmath*}.
Due to the above mentioned symmetry of the system boundaries, the third integral can be rewritten to match the first one, so that
\begin{dmath*}
    \overline{r^{2}} = \frac{1}{L} \left(2\int_{0}^{R}\overline{{r}_{b}^2}(x) \dif x + \int_{R}^{L-R}\overline{{r}_{f}^2} \dif x \right)
\end{dmath*}.
And with \eqref{r-squared-f} and \eqref{r-squared-b}, the final expression for $R \in \left[0,\frac{L}{2}\right)$ becomes
\begin{dmath*}
    \overline{r^{2}} = \frac{1}{L} \left(2\int_{0}^{R}\frac{R^{3}+x^{3}}{3(x+R)} \dif x + \int_{R}^{L-R}\frac{R^{2}}{3} \dif x \right)
    = \frac{1}{L}\left(\frac{5}{9}R^{3}+\frac{1}{3}(L-2R)R^{2}\right) \\
    = -\frac{1}{9}\frac{R^{3}}{L}+\frac{1}{3}R^{2}
\end{dmath*}.

\subparagraph{2. Case $\bm{R \in \left[\frac{L}{2},L\right]}$:} This case needs an extra calculation for the average squared distance $\overline{r_{db}^{2}}(s)$ of a point $r$ in the double truncated interaction range $[-s,L-s] \subset [-R,R]$ (e.g. one ant) to the original intervals center at $0$ (e.g. the other ant), when the interaction range overlaps with both system boundaries. \\
The calculation of $\overline{r_{db}^{2}}(s)$, similar to those of $\overline{r_{f}^{2}}$ and $\overline{r_{b}^{2}}(s)$, is
\begin{align}
\label{eq:r-squared-db}
    \overline{r_{db}^{2}}(s) = \frac{1}{L} \int_{-s}^{L-s} r^{2} \dif r = \frac{(L-s)^{3}+s^{3}}{3L}\, .
\end{align}
In order to also calculate the system wide average over these average squared distances for the case of $R \in [\frac{L}{2},L]$, the system interval $[0,L]$ again needs to be split up into three regions (left, middle and right). The only differences are that the average squared distances calculated in the middle region now result from interaction ranges overlapping with both system boundaries, and a change in the region boundaries. To clarify, the three parts are:
\begin{enumerate}
    \item The average squared distances resulting from interaction ranges overlapping with the left system boundary: $\frac{1}{L-R}\int_{0}^{L-R}\overline{{r}_{b}^2}(x) \dif x\, $.
    \item The average squared distances resulting from interaction ranges overlapping with both system boundaries: $\frac{1}{2R-L}\int_{L-R}^{R}\overline{r_{db}^{2}}(x) \dif x\, $.
    \item The average squared distances resulting from interaction ranges overlapping with the right system boundary: $\frac{1}{L-R}\int_{R}^{L}\overline{{r}_{b}^2}(L-x) \dif x\, $.
\end{enumerate}
Again, using a weighted average, the analogous system wide average is
\begin{dmath*}
    \overline{r^{2}} = \frac{1}{L} \left(\int_{0}^{L-R}\overline{{r}_{b}^2}(x) \dif x + \int_{L-R}^{R}\overline{r_{db}^{2}}(x) \dif x + \int_{R}^{L}\overline{{r}_{b}^2}(L-x) \dif x \right) 
    = \frac{1}{L} \left( 2\int_{0}^{L-R}\overline{{r}_{b}^2}(x) \dif x + \int_{L-R}^{R}\overline{r_{db}^{2}}(x) \dif x \right)
\end{dmath*}
And with \eqref{r-squared-b} and \eqref{r-squared-db}, the final expression for $R \in \left[\frac{L}{2},L\right]$ becomes
\begin{dmath*}
    \overline{r^{2}} = \frac{1}{L} \left(2\int_{0}^{L-R}\frac{R^{2}+x^{2}}{2(x+R)} \dif x + \int_{L-R}^{R}\frac{(L-x)^{2}+x^{2}}{2L} \dif x \right)
    = \frac{1}{18}L^{2}-\frac{1}{3}LR+R^{2}-\frac{5}{9}\frac{R^{3}}{L}
\end{dmath*}.
\ \\
Putting both cases together in non-dimensional terms ($\lambda=\frac{R}{L}$) finally gives the expression \eqref{interaction-radius}.


%

%
%
\end{document}